\documentclass[10pt,letterpaper]{article}
\usepackage[top=0.85in,left=2.75in,footskip=0.75in]{geometry}

\usepackage{amsmath,amssymb}

\usepackage{changepage}

\usepackage[utf8x]{inputenc}

\usepackage{textcomp,marvosym}

\usepackage{cite}

\usepackage{nameref,hyperref}

\usepackage[right]{lineno}

\usepackage{microtype}
\DisableLigatures[f]{encoding = *, family = * }

\usepackage[table]{xcolor}

\usepackage{array}

\newcolumntype{+}{!{\vrule width 2pt}}

\newlength\savedwidth

\raggedright
\usepackage[aboveskip=1pt,labelfont=bf,labelsep=period,justification=raggedright,singlelinecheck=off]{caption}

\makeatletter
\renewcommand{\@biblabel}[1]{\quad#1.}
\makeatother

\usepackage{lastpage,fancyhdr,graphicx}
\usepackage{epstopdf}
\fancyheadoffset[L]{2.25in}
\fancyfootoffset[L]{2.25in}
\newcommand{\br}[1]{\left<#1\right>}

\newcommand{\celegans}[0]{{\it C.~elegans}}

\newcommand{\sethi}[0]{\{h_i\}}

\newcommand{\setJij}[0]{\{J_{ij}\}}
\newcommand{\si}[0]{s_i}
\newcommand{\setsi}[0]{\{s_i\}}
\newcommand{\sj}[0]{s_j}
\newcommand{\sm}[0]{s_m}
\newcommand{\st}[0]{s_t}

\newcommand{\FIM}[0]{F_{mt,m't'}}

\begin{document}
\vspace*{0.2in}

\begin{flushleft}
{\Large
\textbf\newline{Discovering sparse control strategies in \celegans} 
}
\newline
\\
Edward D.~Lee\textsuperscript{1*},
Xiaowen Chen\textsuperscript{2},
Bryan C.~Daniels\textsuperscript{3}
\\
\bigskip
\textbf{1} Complexity Science Hub Vienna, Josefstaedterstrasse 39, Vienna 1080, Austria
\\
\textbf{2} Laboratoire de physique de l'\'Ecole normale sup\'erieure, CNRS, PSL Universit\'e, Sorbonne Universit\'e, Universit\'e de Paris, 75005 Paris, France 
\\
\textbf{3} School of Complex Adaptive Systems, Arizona State University, Tempe, AZ 85287, USA
\\
\bigskip

%
%
E.D.L.~designed the research. E.D.L.~and X.C.~wrote the code. All authors analyzed the results and edited the manuscript.





* edlee@csh.ac.at

\end{flushleft}
\section*{Abstract}
Biological circuits such as neural or gene regulation networks use internal states to map sensory input to an adaptive repertoire of behavior. Characterizing this mapping is a major challenge for systems biology, and though experiments that probe internal states are developing rapidly, organismal complexity presents a fundamental obstacle given the many possible ways internal states could map to behavior. Using \celegans\ as an example, we propose a protocol for systematic perturbation of neural states that limits experimental complexity but still characterizes collective aspects of the neural-behavioral map. We consider experimentally motivated small perturbations --- ones that are most likely to preserve natural dynamics and are closer to internal control mechanisms --- to neural states and their impact on collective neural behavior. Then, we connect such perturbations to the local information geometry of collective statistics, which can be fully characterized using pairwise perturbations. Applying the protocol to a minimal model of \celegans\ neural activity, we find that collective neural statistics are most sensitive to a few principal perturbative modes. Dominant eigenvalues decay initially as a power law, unveiling a hierarchy that arises from variation in individual neural activity and pairwise interactions. Highest-ranking modes tend to be dominated by a few, ``pivotal'' neurons that account for most of the system's sensitivity, suggesting a sparse mechanism for  control of collective behavior.

\section*{Author summary}
The relationship between underlying biological circuitry and behavior is complex and difficult to probe experimentally. Part of the problem is that, for organisms of even modest size, there are an overwhelming number of possible combinations of interacting circuit components. We develop a theoretical framework to simplify this problem with experiments that change the system minimally with small perturbations. In the realm of small perturbations, not only does system behavior remain close to normal but the range of possible perturbations is greatly reduced to only pairs of experimental targets. We demonstrate such a perturbation using a minimal model of neural activity in the \celegans\ worm. We find that a few combinations of ``pivotal'' neurons strongly affect the statistics of synchronous activity, suggesting they may be important for neural control of behavior. Our work suggests feasible, perturbative experiments to map how the physical components of an organism control emergent collective activity.


\section*{Introduction}

Control of complex dynamical networks is a problem of major interest in biology for finding drivers of diseased states \cite{gohHumanDisease2007,vidalInteractomeNetworks2011,zhangIntegratedSystems2013} and determinants of behavior \cite{scholzPredictingNatural2018,moroneSymmetryGroup2019,liuControllabilityComplex2011,tangIdentifyingControlling2012}. In neural systems, the question of which and how many neurons correspond to controllability is largely open.  While there is evidence that single-neuron manipulation is sufficient to induce behavioral change in some cases  \cite{houwelingBehaviouralReport2008, huberSparseOptical2008}, other research shows behavioral information to be encoded amongst many neurons \cite{katoGlobalDynamicsCelegans2015, susoyCelegansMating2020, scholzPredictingNatural2018}. Identifying interesting neurons to probe experimentally and choosing how to perturb them is difficult: in principle, a thorough experiment would require a combinatorially large number of procedures to test all the possible ways that neural targets could be modified. Theoretical tools provide a way of winnowing down the number using control-theoretic analysis of dynamical systems models \cite{yanNetworkControl2017,BorDan21}, structural analysis \cite{moroneFibrationSymmetries2020,moroneSymmetryGroup2019,ZanYanAlb17}, and network properties \cite{delferraroFindingInfluential2018}, amongst other approaches \cite{lynnPhysicsBrain2019}. Here, we develop a theoretical framework that explicitly considers perturbation experiments to discover candidate control neurons. Our framework suggests a way of leveraging recent developments in optogenetics \cite{boydenMillisecondTimescale2005, mardinlyPreciseMultimodal2018} that allow fast, precise control of specific neruons and even the regulation of brain states with the closed-loop optogenetic \cite{grosenickClosedLoop2015, zrennerClosedLoop2016,kimIntegrationOptogenetics2017,pokalaInducibleTitratable2014} or optoelectronic systems \cite{gutrufFullyImplantable2018}.

\begin{figure}\centering
	\includegraphics[width=.7\linewidth]{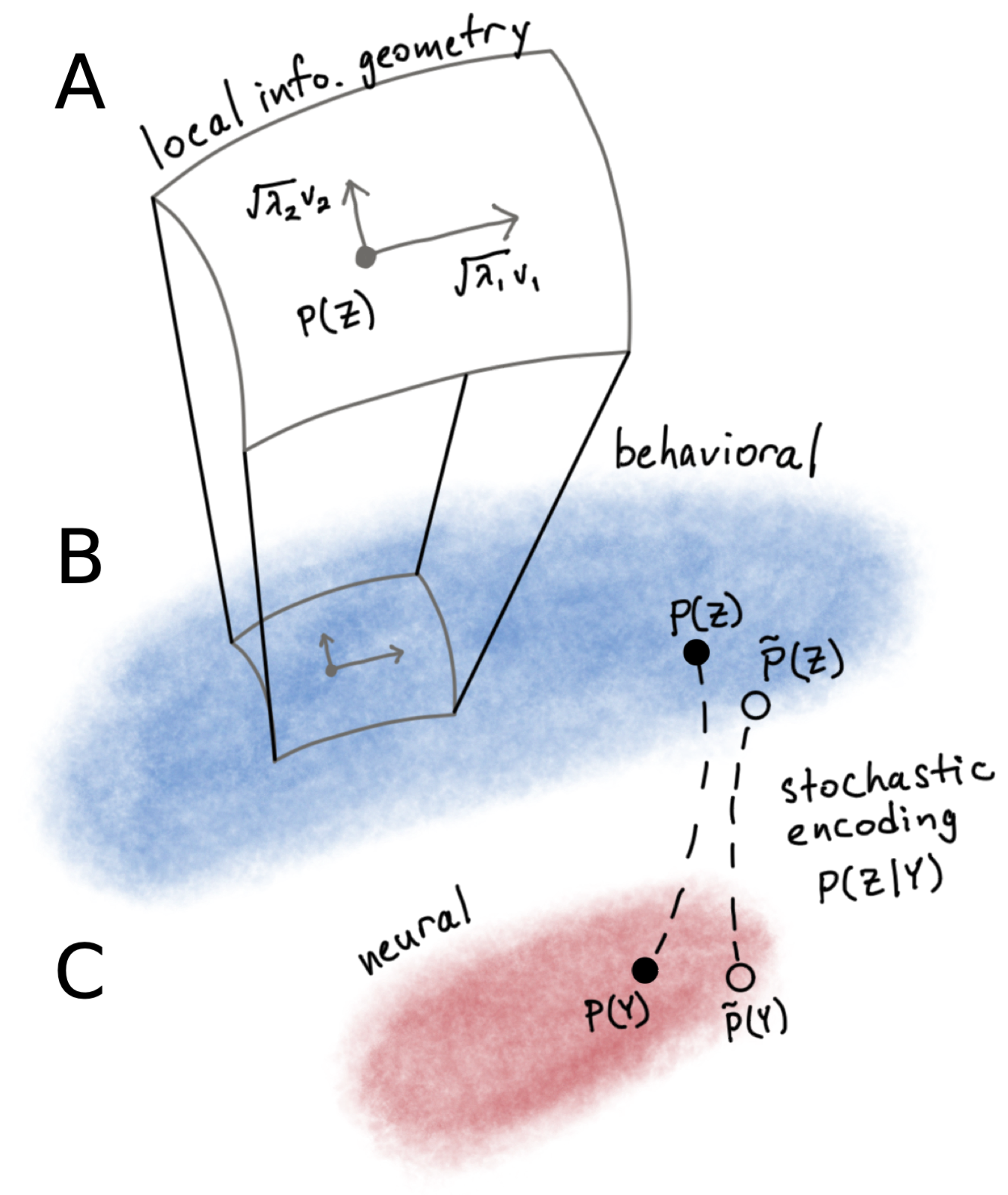}
	\caption{(A) Local curvature of a manifold of probability distributions $P(Z)$ is captured by the Fisher information matrix (FIM). FIM eigenvectors $v_i$ represent axes of combinations of neural perturbations. Eigenvalues $\lambda_i$ denote strength of curvature, or inverse sensitivity. (B) Set of possible behavioral distributions $P(Z)$ as a high-dimensional manifold.  (C) Neural behavior represented as ensemble $P(Y)$ over states of activity $Y$. Relationship between $P(Y)$ and behavioral distribution $P(Z)$ is given by stochastic encoding $P(Z|Y)$ that maps neural activity to behavior.}\label{gr:info geometry}
\end{figure}

The question of control is in essence about how neural states map to behavior. When the mapping is not deterministic, this is a problem of stochastic encoding of the behavioral state $Z$ in the global neural state $Y$. 
Whether from noise, experimental limitations, or other sources of uncertainty, there is for a given neural state $Y$ a multiplicity of $Z$ characterized by the conditional probability distribution $P(Z|Y)$. In practice, we cannot specify a unique neural state $Y$ due to experimental uncertainty or limits to measurement. Instead, we measure some distribution over configurations $P(Y)$ that we then map to some probability distribution of behavioral outcomes $P(Z)$. By considering the realm of possible distinct neural configurations, we obtain the stochastic mapping formulation in Figure~\ref{gr:info geometry}, where at each point we specify a distribution of neural activity, of behavior, and the relationship between the two that accounts for imprecision and uncertainty.

A classic formulation of stochastic mapping in sensory encoding is that of individual neurons that encode increasingly complex patterns of sensory stimuli as information is funneled up a hierarchy \cite{barlowSingleUnits1972,quirogaSparseNot2007}. In this case, the mapping between $Y$ and $Z$ is precise. For example, we take $Y$ to be the firing rate of some neuron and $Z$ to be the similarity of the stimulus to one's grandmother. Since firing rate is a mean of a noisy measurement, it is more appropriate to interpret the average firing rate to characterize a distribution $P(Y)$, e.g.~a Poisson distribution for the number of times a neuron fires within a time frame. Then, varying the mean activity level of a single neuron traces out a curve in the space of possible distributions $P(Y)$. In the space of possible perceptions $P(Z)$, representing for example a probability distribution over recognizing ``grandma'' or not and given by $P(Z) = \sum_Y P(Z|Y)P(Y)$, we likewise trace out a curve as a single neural parameter is varied.

Later paradigms generalized the focus on single neurons to a sparse set of neurons that together span the range of possible sensory inputs. In the formulation of sparse coding, where a basis set of neurons are extracted from measuring neural response to a set of visual stimuli, leading to abstract image features such as edges, corners, and contours \cite{danielsSparseCode2012,olshausenSparseCoding1997}. The idea of sparse coding is that typical visual scenes can be represented with activity in only a few neurons, with each neuron representing a separate high-level feature. Then, it is the joint activity amongst these distributed components that captures the whole image, an idea with multiple variations in the literature \cite{spanneQuestioningRole2015,beyelerNeuralCorrelates2019}.

Despite the generalization away from single neurons in the study of sensory encoding \cite{oizumiInformationLoss2011}, an echo of it still rings in experimental studies of neural control of behavior. Some experiments rely on the notion of individual (or sets of similar) ``driver'' neurons that --- when ablated, silenced, or otherwise modified --- substantially change behavioral outcomes \cite{grayNavigationCelegans2005,ikedaContextdependentOperation2020,ouelletteGateandSwitchModel2018}. While some individual cells, like interneurons connecting distal parts of the body, are essential for normal behavior, some higher-order behaviors depend on many neurons in a way that is robust to the function of individual cells \cite{changHypoxiaHIF12008,lashleyBrainMechanisms1929,damasloReturnPhineas1994}. Indeed, the finding that some neural circuits encode sensory input in a sparse, distributed way raises the question of whether neural control of behavior also follows similar principles \cite{quirogaSparseNot2007,stephensDimensionalityDynamics2008,scholzPredictingNatural2018,spanneQuestioningRole2015}.

\begin{figure}\centering
	\includegraphics[width=.75\linewidth]{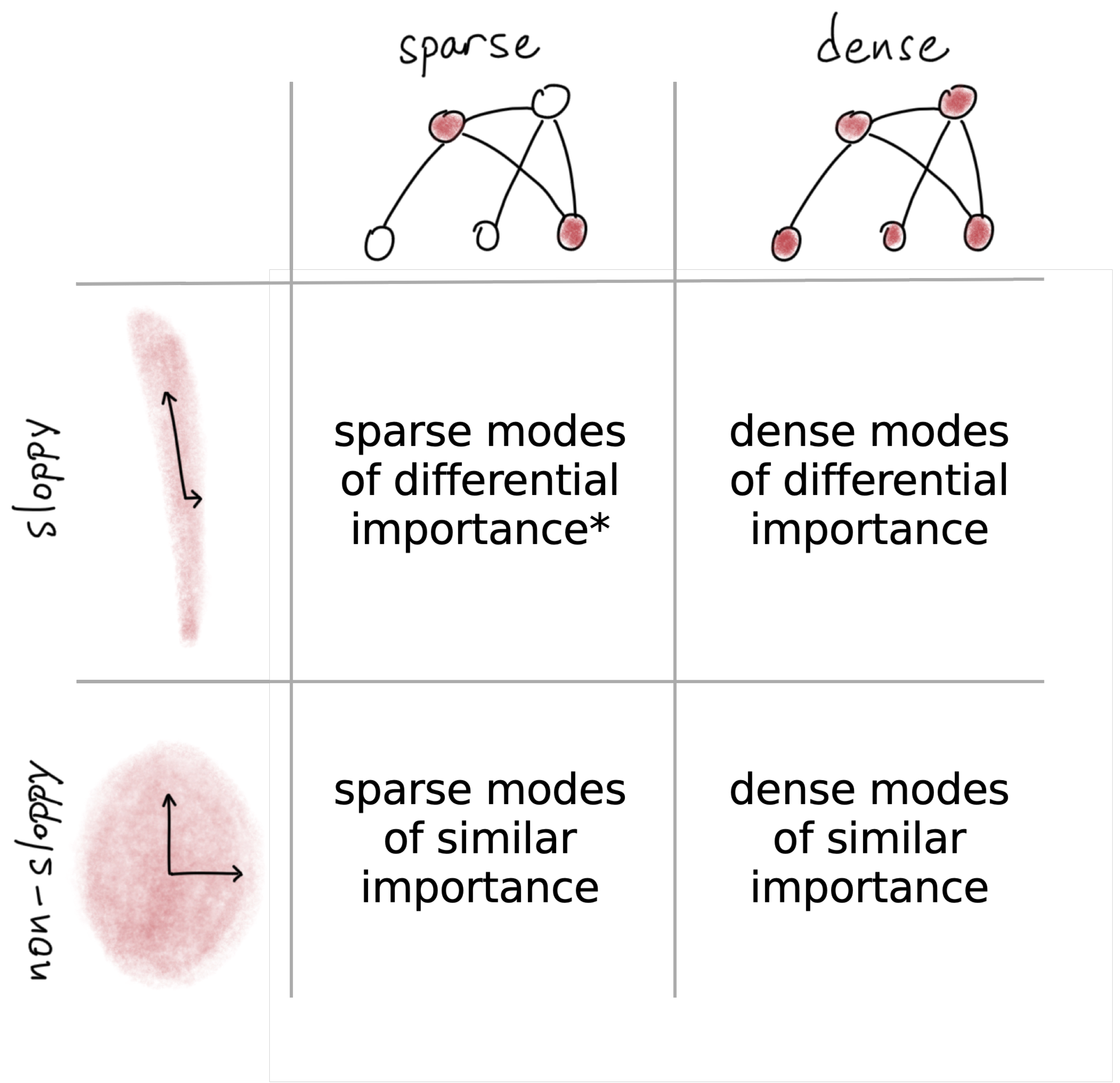}
	\caption{Four possibilities for the sparsity of collective modes of sensitivity. When collective activity shows sloppy structure, the local information geometry is elongated, with both stiff sensitive directions and sloppy insensitive ones. With dense collective encoding, multiple components each matter equally. The starred combination, with sloppy, sparse combinations of neurons, aligns with centralized control.}\label{gr:ig examples}
\end{figure}

Here, we explore two potential sources of sparsity in neural control of behavior. First, a desired output could be equally well produced by many possible changes, or modes, at the neural level or could be much more easily produced by only one particular change.  Second, each mode can itself be sparse or dense in the number of neurons involved \cite{schneidmanWeakPairwise2006,hopfieldNeuralNetworks1982}. These possibilities lead to the four hypotheses we draw in Figure~\ref{gr:ig examples}, where control is sloppy or non-sloppy with either sparse or dense collective modes of sensitivity. 

In the context of distributions of neural activity, variation in sensitivity of neural activity is described by the local dependence of the neural activity distribution $P(Y)$ on parameters specifying its form. As pictured in the top of Figure~\ref{gr:info geometry}, ``stiff'' directions correspond to large changes in $P(Y)$ being caused by small changes to neural parameters \cite{transtrumModelReduction2014,transtrumPerspectiveSloppiness2015}. In the ``sloppy'' directions, changes to neural parameters are ineffectual, and only dramatic perturbations cause a comparable change in $P(Y)$. The relative elongation of the local information geometry disappears when encoding places no special importance on any neural subgroup such that each plays a commensurate role, as in some forms of population coding \cite{maunsellFunctionalProperties1983,wuPopulationCoding2002}. In this way, the local information geometry informs us about the first notion of sparsity mentioned above.  Additionally, the second notion of sparsity is captured by the degree to which stiff directions involve changes to only a few neurons. Of particular interest is the combination of sparse and sloppy structure, where collective sensitivity would be concentrated in only a few perturbative sets that each include only a few neurons. Such a possibility reflects centralized control, which would seem to be optimal for simplifying the multiplicity and complexity of control nodes. If real world examples display a range of structure depending on organism and function, we need a flexible methodology for inferring such variation from data.

We propose a perturbative, experimental approach to study stochastic encoding between internal states and output behavior.  Perturbing the statistics of neural activity parameterizes motion in the space of all possible distributions, the rate of which quantifies sensitivity to neural changes. Our focus on small perturbations allows us to derive a local approximation that is the simplest complete description of local sensitivity.

For concreteness, we focus on the model animal \textit{Caenorhabdits elegans}. With only 302 neurons and structural connectome mapped, \celegans\ presents a particularly interesting example with which to explore neural control. Current experiments measure whole-brain activity in a freely moving worm and soon will allow matching those neurons to the structural connectome \cite{nguyenWholeBrain2016,venkatachalamPanneuronalImaging2016}. Combined with tools for manipulating neural states \cite{stirmanRealTime2011}, our proposed experimental protocol could be used to search for neural centers important to collective activity. As an example of the proposed method, we start with minimal models matching the statistics of neural activity in \celegans, and then compute the local information geometry to extract combinations of neurons that, if perturbed, would strongly affect collective statistics. Though we focus on the neural activity of \celegans, our procedure could be adapted to other measures of internal states such as hormone concentration, gene expression, or other experimentally accessible quantities, and to output behavior including morphological descriptors such as orientation, velocity, extension, or a stereotyped behavioral sequence \cite{bermanPredictabilityHierarchy2016}. In this sense, our work provides a generalizable theoretical framework with specific measurements and predictions that can be tested experimentally in \celegans\ and other biological systems.

\section*{A perturbative thought experiment}
An exhaustive protocol for probing the \celegans\ neural system would be to test simultaneous perturbations to subsets of neurons in ways that are sympathetic, antagonistic, and of varying magnitude. Unfortunately, the combinatorial explosion in the number of possibilities makes such an experiment impossible.
In the limit of small perturbations, however, all possible effects can be extracted from  observing only the response of the system to perturbations of pairs of neurons. We use this limit in part because pairwise relationships between perturbations becomes a sufficient description of system response. Additionally, it is often the case that collective statistics of neural activity can be captured with pairwise neural statistics without needing to invoke higher-order correlations \cite{schneidmanWeakPairwise2006,merchanSufficiencyPairwise2016,chenSearchingCollective2019}, so we might anticipate that the pairwise approximation could be a reasonable one even for larger changes.  We demonstrate how this simplification could be useful by carrying out a thought experiment that mimics the perturbations {\it in silico}, making use of a minimal model of neural activity.

To formalize the set of perturbations, we consider the discretized gradient of calcium concentration $\si$ of neuron $i$, which is either decreasing $\si=-1$, flat $\si=0$, or increasing $\si=1$ with probabilities $p(s)$ of observing the set $s\equiv\setsi$. Such discretization has been explored in previous work \cite{katoGlobalDynamicsCelegans2015}. Then, a perturbation to underlying neural activity is reflected in a modified probability distribution over discrete states $\tilde{p}(s) = p(s)+\Delta(s)$. A unique measure of distinguishability, the information divergence, between the original distribution and the modified one reduces to 
\begin{align}
	D_{\rm KL}[p||\tilde p] &\approx \frac{1}{2} \sum_{s} \frac{\Delta(s)^2}{p(s)} = \frac{1}{2} \sum_s \sum_{ij} F_{ij} dv_i(s) dv_j(s).\label{eq:simple fim}
\end{align}
Then, the complete set of perturbations is given by the Hessian $F_{ij}$ describing the local sensitivity of $p(s)$ to perturbations along vectors $dv$, which may be constrained by what is accessible in the experiment or model. Because moving from one basis to another is a linear operation, the particular form of the perturbations are not as important as the fact that the set should span the possible set of perturbations. By exploiting this property of analyticity, we can in principle reduce the complexity of the problem by measuring a single set of experimental perturbations.

In a hypothetical experiment, we might effect perturbations by inserting electrodes into immobilized worms --- or, in a more elegant protocol using optogenetic tools, by clamping membrane potential and effectively coupling neurons to one another via an external circuit to control the strength and timing of perturbation. Here, we use a  perturbation that with small probability copies the state of one neuron to another, chosen to resemble increase or decrease in synaptic strength.  Specifically, we select pairs, identifying a ``target'' neuron in state $\st$ and a ``matcher'' neuron in state $\sm$, and clamp the matcher to the target's state, $\sm=\st$, with small probability $\epsilon$ (see Appendix~\ref{si sec:experiment} for more details). Compared to more common clamping and ablation techniques \cite{pokalaInducibleTitratable2014,xuHighlyEfficient2016}, this proposed closed-loop perturbation is not as drastic, mostly preserves natural dynamics, and is closer to internal control mechanisms such as synaptic modulation that do not require constant input from the outside \cite{sukClosedLoop2017, newmanOptoFeedback2015, prinzDynamicClamp2004}.

\section*{Maximum entropy (maxent) model example}
We build a minimal model of the collective statistics of \celegans\ on which to run such a protocol {\it in silico}, systematically perturbing it and extracting from its response the sensitive modes of collective neural statistics. The two examples we consider are of $N=50$ neuron subsets from the anterior neural network of the immobilized worm \cite{scholzPredictingNatural2018}. With a discretized time series representing the three possible configurations for neuron $m$, we calculate time-averaged probabilities of being in state $k$,
\begin{align}
	r_k(\sm) = \frac{1}{T}\sum_{t=1}^T\delta_{\sm,k}(t),
\end{align}
where the Kronecker delta function $\delta_{s_m,k}$ indicates when the state of the neuron $s_m$ is $k$ at time $t$ over the duration of the experiment $T$. Similarly, the pairwise probabilities of agreement between any two neurons is $\sum_{t=1}^T\delta_{\sm, \st}(t)/T$ and similar higher-order correlations describe the probabilities of agreement between multiple neurons. These higher-order correlations are not necessarily given by the lower-order statistics, but only accounting for the pairwise correlations is sufficient to capture the higher-order correlations \cite{chenSearchingCollective2019} (see Figures~\ref{si gr:maxent} and \ref{si gr:maxent second}).

An advantage of the maxent approach is that it captures the statistics of neural activity in a way that reflects latent physical interactions while also obeying a quantitative formulation of Occam's Razor \cite{morcosDirectcouplingAnalysis2011,volkovInferringSpecies2009}. The maxent principle ensures that a model of the probability distribution $p(s)$ only matches specified constraints but otherwise remains as structureless as possible. This can be done by maximizing the model's information entropy, $S=-\sum_{s} p(s)\log p(s)$, here a sum over all $3^N$ possible configurations, with the standard method of Langrangian multipliers \cite{coverElementsInformation2006,jaynesInformationTheory1957,leeStatisticalMechanics2015,schneidmanWeakPairwise2006}. When the set of average individual neural activity $\{\br{r_k(s_m)}\}$ is constrained, the maxent procedure gives the ``independent model'' where the probability of finding neuron $i$ in state $k$,
\begin{align}
	r_k(\si) &= e^{h_{i,k}}/Z_i,\label{eq:indpt model}
\end{align}
is determined by the bias $h_{i,k}$, with normalization term $Z_i = \sum_{k=-1}^1 e^{h_{i,k}}$. A larger bias $h_{i,k}$ relative to another state $k'$ indicates that it is more likely to find the neuron in state $k$ than $k'$. This independent model fails to capture correlations in neural activity.

A simple variation that does not assume independence involves constraining the pairwise correlations. Given the finite-size of the data set, we constrain the pairwise coincidence probabilities to obtain the pairwise maxent model with distribution given the energy function $E(s)$,
\begin{align}
\begin{aligned}
	p(s) &= \left.\exp\left[ -E(s) \right] \right/ Z,\\
	Z &= \sum_{s} \exp\left[ -E(s) \right],\\
	E(s) &= -\sum_{i<j}^N J_{ij} \delta_{\si,\sj} - \sum_{k=-1}^1\sum_{m=1}^N \delta_{k,\si}h_{i,k}.
\end{aligned}\label{eq:pair maxent}
\end{align}
The log-probability of configuration $s$ varies with its energy such that lower energy means larger probability. As with the independent model, increasing neural bias $\{h_{i,k}\}$ pushes neurons to state $k$ and increasing the magnitude of couplings $\setJij$ magnifies the tendency of pairs to agree or to disagree when negative. Though the couplings are in principle exactly specified by the pairwise correlations in the data, sampling and experimental noise mean that there are many possible sets of couplings that align within the statistical variation of the data sample. We focus on a numerical solution that recovers topological structure in the coupling network and has been shown to faithfully capture collective neural patterns \cite{chenSearchingCollective2019} (see Appendix~\ref{si sec:solution} for solutions from an approximate gradient descent method).
The solution then defines a minimal interaction network that is distinct from the pattern of pairwise correlations (see Figure~\ref{si gr:maxent}).

\begin{figure}\centering
	\includegraphics[width=.8\linewidth]{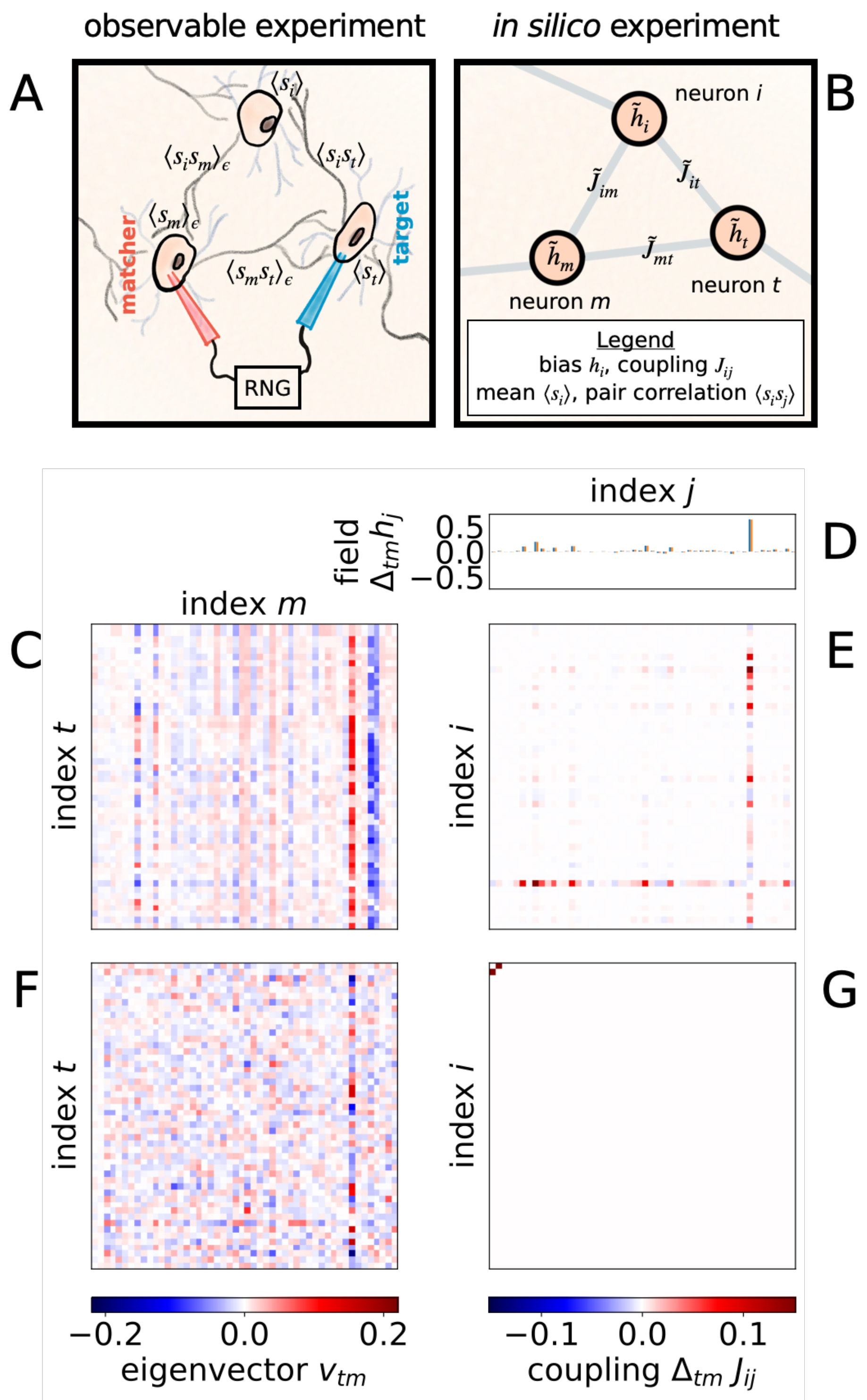}
	\caption{(A) Perturbation thought experiment consists of clamping matcher neuron $m$ to the state of target neuron $t$ with some small probability $\epsilon$ when indicated by a random number generator (RNG). We picture electrodes controlling membrane voltage, but optogenetic protocols may be more elegant. (B) Perturbation corresponds to modifying fields and couplings in a pairwise maxent model. (C) Principal eigenmatrix mapped to change in (D) biases $\sethi$, each direction indicated by a different color, and (E) couplings $\setJij$ for $\epsilon=10^{-4}$. (F) Diffuse perturbation in space of observables using replacement rule from Eq~\ref{eq:replacement rule} corresponds to (G) localized ``natural'' perturbation to one coupling (note the nonzero values in the top left corner). Localized coupling perturbation is nontrivial and nonintuitive in terms of the observable perturbation.}\label{gr:comparison}
\end{figure}

\section*{Mapping perturbations from experiment to simulation}

With the maxent model, we enact our {\it in silico} thought experiment. Continuing with the formulation presented above, we connect a pair of target neuron $t$ and matcher neuron $m$ such that $\sm$ is clamped to match $\st$ with small probability $\epsilon\ll 1$. Then, the modified probability $\tilde r_k$ that the matcher neuron is in state $k$ is the mixture
\begin{align}
	\tilde r_k(\sm) &\equiv (1-\epsilon) r_k(\sm) + \epsilon\, r_k(\st).\label{eq:replacement rule}
\end{align}
As a result of the perturbation in Eq~\ref{eq:replacement rule}, the statistics describing neuron $m$ becomes more like those of neuron $t$. This procedure likewise modifies the matcher neuron's coincidence $p(\sm=s_j)\rightarrow\tilde p(\sm=s_j)$ with all other neurons indexed $j$ as if modifying synaptic weights \cite{leeSensitivityCollective2020}. If we were to take sufficiently large $\epsilon$ in experiment, we would expect such intervention not to remain localized to neuron $s_m$ but to alter the neighbors and eventually the neighbors of neighbors and so on. If $\epsilon$ is small, then we expect a localized perturbation to be able to well-approximate the desired change in statistics given in Eq~\ref{eq:replacement rule}. By taking the limit where the perturbation is sufficiently small and assuming that it then remain localized, we specify a perturbation that corresponds to a unique change of parameters.

Such uniqueness implicitly depends on constraining ourselves to considering perturbations that move us along the family of pairwise maxent models. This constraint is not reflected in the replacement rule in Eq~\ref{eq:replacement rule}, which compatible with an infinite variety of changes to higher-order correlations. In other words, we do not necessarily need to restrict ourselves to considering only the energy function defined in Eq~\ref{eq:pair maxent}, but we could allow for higher-order interactions to appear under perturbation. This introduces ambiguity that can only be resolved by choosing the form of the perturbations. Once we have chosen to fix the structure of the energy function from the maxent principle, however, we do not allow a perturbation to arbitrarily alter the form. This assumption is consistent with the widespread observation that the pairwise model generally captures well collective features of biological neural networks of modest size \cite{schneidmanWeakPairwise2006,tkacikSpinGlass2009,tkacikSearchingCollective2014,meshulamCoarseGraining2019,bartonIsingModels2013,roudiPairwiseMaximum2009}. Thus, we use the pairwise maxent model not only to specify a minimal, compressed representation of neural statistics, but also to specify how the probability distribution evolves under the replacement rule specified in Eq~\ref{eq:replacement rule}.

This replacement rule maps experimental perturbations to observed statistics in contrast with starting with model parameters as is often the first instinct for theorists. In the latter formulation, the fields and couplings would form the basis for perturbations, also known as ``natural'' or ``canonical'' perturbations \cite{amariInformationGeometry2016}. This latter perspective adheres to the physical intuition that the parameters from Eq~\ref{eq:pair maxent} reflect latent physical interactions. When the pairwise maxent model instead serves as an approximation of the underlying physical interactions, a natural perturbation may be mediated by unknown parameters instead of fields and couplings \cite{merchanSufficiencyPairwise2016,bialekRediscoveringPower2007}, and this renders the straightforward model perturbation difficult to translate into a literal experiment.
In comparison, defining perturbations in observable terms as in Eq~\ref{eq:replacement rule} allows for easier experimental interpretation and for consistent comparison of inferred models \cite{leeSensitivityCollective2020}.

If it is collective neural activity that encodes behaviorally relevant information \cite{hopfieldNeuralNetworks1982,schneidmanWeakPairwise2006}, then we can use collective properties as a proxy for behavior. As one example measure of collective activity, we consider the probability that the number of neurons in each of the three states, $\phi_{\rm fine}(n_1,n_2,n_3)$, is described when ordered by $n_1\geq n_2\geq n_3$ such that $n_1$ corresponds to the size of the plurality and $n_3$ the smallest minority (i.e.~there is no fixed correspondence to ``rise,'' ``fall,'' and ``flat''). This presents a measure of synchrony more fine-grained than $\phi_{\rm coarse}(n_1)$ that only considers the number of neurons in the plurality $n$. Crucially, the pairwise maxent model also captures these statistics. While neither statistic differentiates between the orientation of the states, we limit ourselves here to simpler and less computationally expensive statistics of collective neural activity \cite{tkacikSimplestMaximum2013a,bialekBiophysicsSearching2012}, noting that other statistics may easily be incorporated in the way we outline.

Under perturbation of a pair of matcher and target neurons, the synchrony distribution $\phi$ is mapped to an altered $\tilde\phi$ that depends on the strength of perturbation $\epsilon$. To measure the distance between the two distributions, we use the Kullback-Leibler (KL) divergence $D_{\rm KL}$ as in Eq~\ref{eq:simple fim} \cite{amariInformationGeometry2016}. In the limit of infinitesimal perturbation, $\epsilon\rightarrow0$, KL divergence reduces to the Fisher information
\begin{align}
\begin{aligned}
	\FIM &\equiv \lim_{\epsilon\rightarrow0} \frac{2}{\epsilon^2} D_{\rm KL}\left[\phi\Big|\Big|\tilde \phi\right] =  \lim_{\epsilon\rightarrow0} \frac{2}{\epsilon^2} \sum_{n_1\geq n_2\geq n_3} \phi \log \left(\frac{\phi}{\tilde\phi}\right) \equiv  \lim_{\epsilon\rightarrow0} \frac{2}{\epsilon^2} \br{ \log \left(\frac{\phi}{\tilde\phi}\right) } \\
	&=  \sum_{\theta_i,\theta_j}\br{\frac{\partial^2 \log \phi}{\partial\theta_i\theta_j}\mathcal{J}_{mt,m't'}^{ij}},
\end{aligned}\label{eq:fim}
\end{align}
where the Jacobian $\mathcal{J}_{mt,m't'}^{ij}$ transforms the maxent parameters such as fields and couplings $\{\theta_i\}$, here ordered along a single index $i$, into the vector of changes that correspond to the observable perturbations specified in Eq~\ref{eq:replacement rule}. In contrast with Eq~\ref{eq:simple fim}, we consider in Eq~\ref{eq:fim} the impact of perturbation on collective synchrony in Eq~\ref{eq:fim} that accounts for the multiplicity of microscopic configurations belonging to a single coarse-grained state \cite{variance_note}.

The Fisher information matrix (FIM), whose entries are defined in Eq~\ref{eq:fim}, is spanned by eigenvectors that describe orthogonal perturbations of the parameters $\{\theta_i\}$, where modes with large eigenvalues describe perturbations to which collective synchrony is highly sensitive and small eigenvalues represent ones to which the system is insensitive \cite{leeSensitivityCollective2020,transtrumPerspectiveSloppiness2015}. As we show in Eq~\ref{eq:fim}, this basis describes the local curvature of a Riemannian manifold generated by information distance. Importantly, the basis vectors can involve a mix of antagonistic and sympathetic perturbations of neurons of varying magnitude, one that would be nontrivial to extract {\it a priori}. Thus, the FIM encodes how quickly coarse-grained configurations $\phi$ change as we modify the system on a microscopic level by perturbing pairs of neurons at a time, with perturbations that mimic changes to synaptic connectivity.

\section*{Leading FIM eigenvalues show Zipfian decay}
\begin{figure*}\centering
	\includegraphics[width=.9\linewidth]{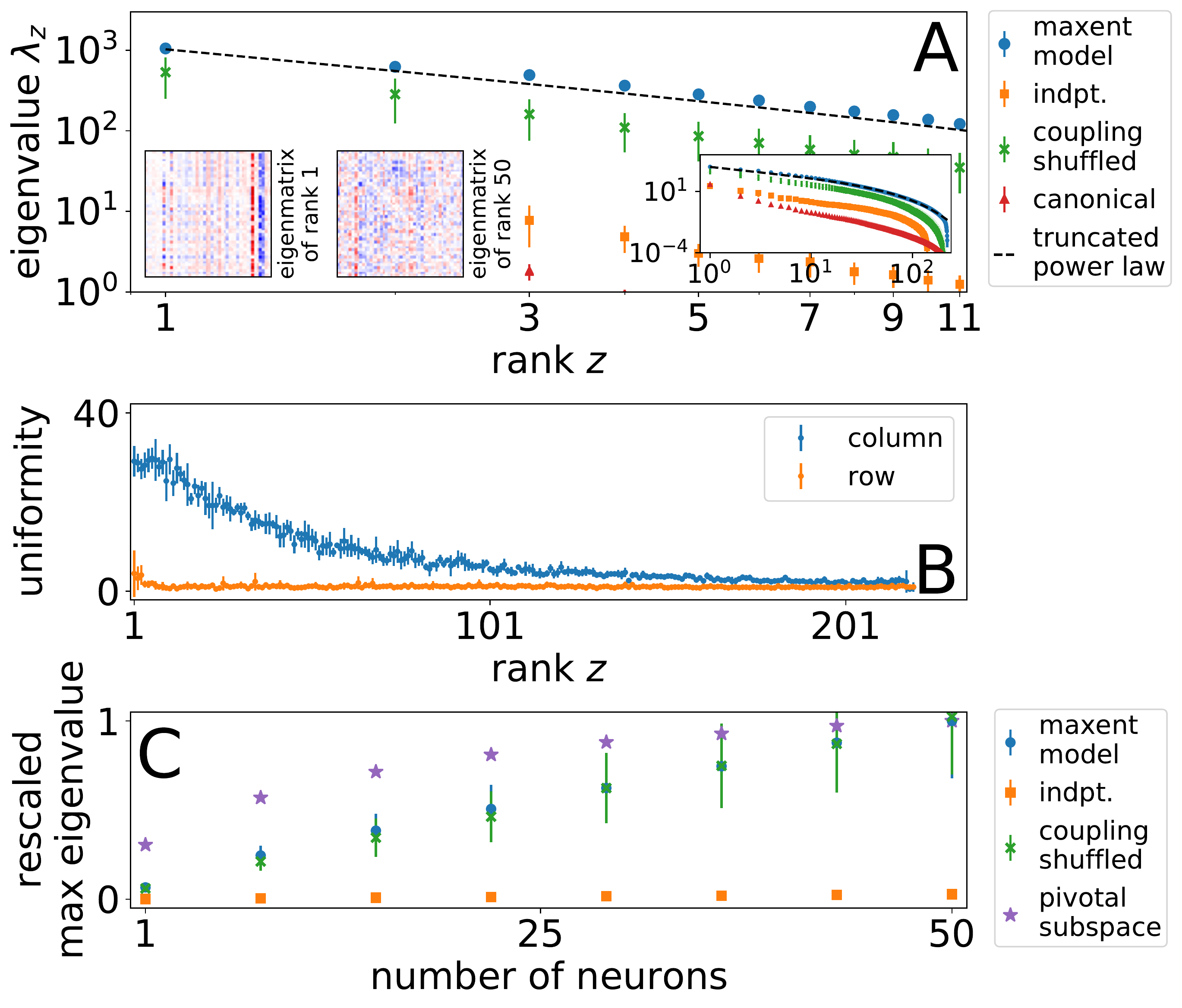}
	\caption{(A) Eigenvalue spectrum of FIM for pairwise maxent model, independent model (indpt.), and shuffled couplings null, with observable perturbations from Eq~\ref{eq:replacement rule} averaged over $M$ Monte Carlo samples ($M=10$ for maxent model, $M=10$ over random shuffles for coupling shuffled, $M=4$ for independent and canonical). For comparison, we show response to canonical perturbation to couplings. Insets on left show example eigenmatrices of rank 1 and rank 50. Inset on right shows full eigenvalue spectrum. Error bars show standard error of the mean. (B) Uniformity from Eq~\ref{eq:uniformity} averaged over samples for eigenmatrices. Error bars represent standard deviation over rank-ordered plots. (C) Principal eigenvalues averaged over random subsets of neurons as a function of subset size rescaled by the average maximum eigenvalue for the maxent model. Sensitivity tends to be much lower for the independent model but of similar scale for shuffled couplings. Error bars represent standard deviation taken over subsets for each permutation or model. Compare with Figure~\ref{si gr:eigenspectrum}.}\label{gr:eigenspectrum}
\end{figure*}

In Figure~\ref{gr:eigenspectrum}A, we show the rank-ordered eigenvalue spectrum of the FIM for collective synchrony $\phi_{\rm fine}$ calculated with the pairwise model in comparison with several null models: independent neurons, pairwise maxent model with couplings randomly shuffled between all pairs of neurons (which preserves the distribution of couplings but not the topology of the interaction network), and canonical perturbations directly modifying each coupling at a time by a fixed amount (Figure~\ref{si gr:exponent fits}). Across all cases, we expect a hard cutoff at the dimensionality of the synchrony distribution $\phi_{\rm fine}$ corresponding to rank $Z_{\rm max}=234$, which is much smaller than a single dimension of the matrix $N^2-N={2{,}450}$ ($Z_{\rm max}=33$ for the coarse-grained collective statistic $\phi_{\rm coarse}$ as we show in Figure~\ref{si gr:eigenspectrum coarse 170419}). In contrast with the others, the independent neuron model has short maximum cutoff well below the theoretical maximum, reflecting the essential role of interactions in mediating pairwise perturbations. The cutoff for the pairwise maxent model reveals the replacement rule in Eq~\ref{eq:replacement rule} sufficient to very nearly span the full dimensionality of synchrony space. This observation confirms that the pairwise model with the pairwise replacement rule generates a useful basis to explore collective sensitivity. 

The rank-ordered spectrum of eigenvalues $\lambda_z$ of rank $z$ initially decays such that each successive level of perturbation returns multiplicatively smaller response, following roughly on a very limited range Zipf's law, $\lambda_z\propto z^{-1}$. In contrast, a simple exponential decay would indicate a sharp cutoff for sensitive modes beyond some rank. We find that exponential decay alone does not describe the  eigenvalue spectrum, and instead find a reasonable fit using a power law with an exponential tail:
\begin{align}
	\lambda_z &= A z^{-\alpha} e^{-z/\bar{z}};\qquad z\leq Z_{\rm max} \label{eq:eig decay}
\end{align}
In Eq~\ref{eq:eig decay}, we have parameters for vertical scaling $A$, exponent $\alpha$, tail $\bar{z}$, and hard cutoff $Z_{\-\rm max}$ beyond which there are no additional eigenvalues or numerical precision errors may be important (see Section~\ref{si sec:fim analysis} for more details about the fitting procedure). For random subsets of $N=50$ neurons considered, we find that this model usually presents a compelling fit and that the scale of the cutoff $\bar{z}\gtrsim30$ indicates a power-law-like regime for the dominant eigenvalues (Figures~\ref{si gr:spectra fit} and \ref{si gr:spectra fit 2}).

Such a scaling regime seems to be a feature of the system statistics, consistently appearing across the null models. The shuffled null model agrees closely with our thought experiment, in which eigenvalue decay tends to be slowest, showing a power-law exponent closest to $\alpha=1$ (see Figure~\ref{si gr:exponent fits}). Perhaps unsurprisingly, the spectra are also scaled to higher sensitivities than the independent model. Though the latter is roughly commensurate in overall sensitivities with canonical perturbations, we might expect perturbations in the space of observable and model parameters to be scaled differently from the transformation of variables (Appendix~\ref{si sec:obs and can}). Thus, this statistical hierarchy reflects the role of component disorder common across the models whose overall collective sensitivity is magnified by interactions.

\section*{FIM eigenvectors reveal pivotal neurons}
Inspecting the corresponding eigenvectors, we find some modes are dominated by perturbations focused only on a few neurons. To better represent the connection between pairwise perturbations and eigenvectors, we reshape eigenvectors into {\it eigenmatrices} $V_{mt}$ such that the elements in a column indicate how matcher neuron $m$ should imitate its target neighbors $t$ in turn. When put into this representation, we often find vertical striations as in the inset of Figure~\ref{gr:eigenspectrum}A. These striations are visible because they are almost exclusively of the same sign, indicating that the mode describes perturbations localized to a single neuron that tend to increase or decrease its correlation with all neighbors simultaneously. 
In contrast, horizontal striations that represent uniform perturbations across all the neighbors of a particular neuron, a kind of global perturbation, tend to be sparser and weaker on average. This emergent pattern contained in the block structure of the FIM suggests that localized, uniform enhancement or suppression of synaptic connections leading to a small set of {\it pivotal} neurons may serve as effective mechanisms for modulating collective activity.

As a more direct analysis of such key neurons, we limit our analysis to perturbations focused on a single matcher neuron at a time and put aside perturbations combining different matcher neurons. When the FIM is ordered first by matchers and then targets for each matcher, these perturbations correspond to blocks along the diagonal of the FIM of dimension $(N-1)\times(N-1)$ for fixed $m$ and variable $t$. From the principal eigenvalues of the diagonal blocks, we find that the single, pivotal neurons with the largest eigenvalues tend to coincide with the ones that manifest in vertical striations of the full FIM (Appendix~\ref{si sec:fim analysis}). These striations correspond to columns with high uniformity. As a measure of this, we define row and column uniformity, respectively, as
\begin{align}
\begin{aligned}
	U_{i} &\equiv \left(\sum_j v_{ij}\right)^2; \qquad V_{j} &\equiv \left(\sum_i v_{ij}\right)^2,
\end{aligned}\label{eq:uniformity}
\end{align}
for each normalized eigenmatrix $v_{ij}$. When we consider the subspace of leading pivotal neurons and compare them with the subspace of randomly chosen neurons, the principal eigenvalues of the former set saturate much more quickly to the leading eigenvalue of the full matrix even when we have only considered about half of the available neurons (Figure~\ref{gr:eigenspectrum}C), indicating that there is a subset that in combination overwhelmingly determines collective sensitivity. This concentration of sensitive modes in a few neurons is an indication of centralized structure that emerges from neural heterogeneity.

\begin{figure}\centering
	\includegraphics[width=.9\linewidth]{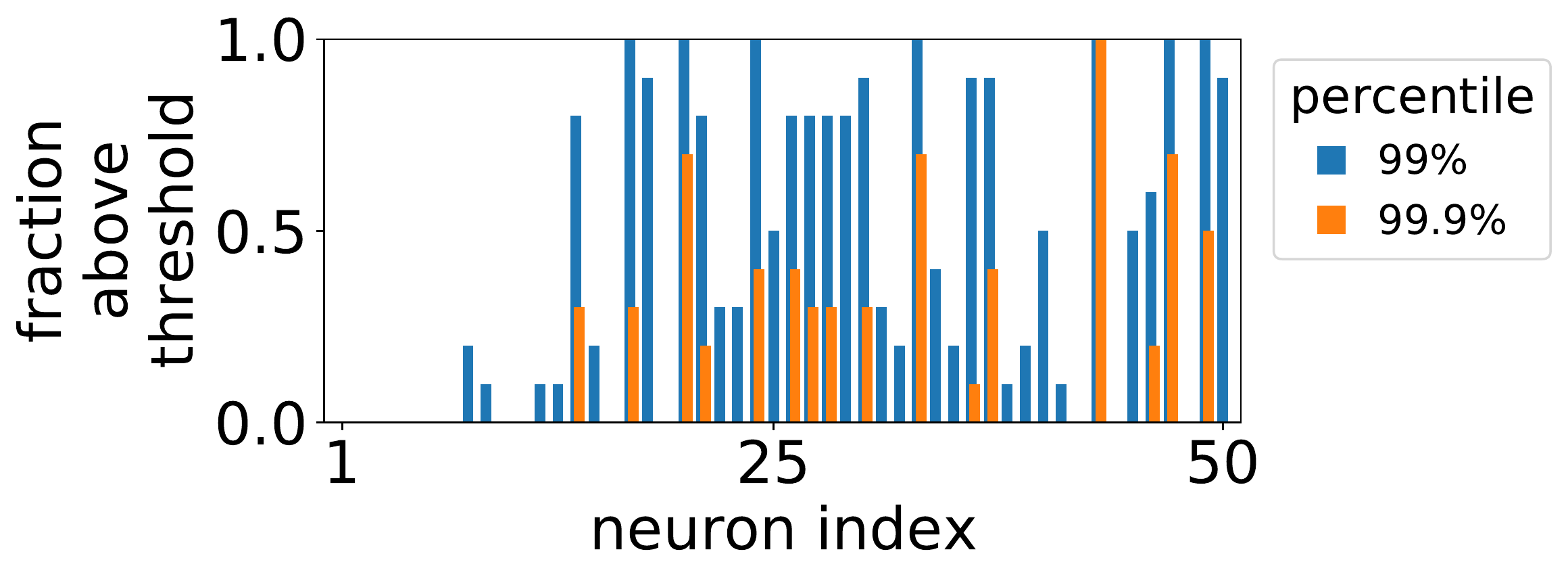}
	\caption{Fraction of times out of 10 Monte Carlo samples that neuron column uniformity is above 99\% and 99.9\% percentile cutoffs. 
	Distribution is column uniformities over all neurons over all eigenmatrices for a single Monte Carlo sample. See also Figure~\ref{si gr:piv neurons}.}\label{gr:piv neurons}
\end{figure}

The patterns that we note both in the eigenvalue spectrum and eigenvector basis hold generally for random neuron subsets of $N=50$ sampled from each experiment, about the maximum possible number that can be sampled reliably \cite{chenSearchingCollective2019}. In contrast with the typical notion that the identities of particular neurons are essential, pivotal neurons fluctuate between subsets and random samples as shown in Figure~\ref{gr:piv neurons}. While some neurons are more frequently identified in spite of random fluctuations for the same subset of $N=50$ neurons, consistently identified pivotal neurons are less pronounced once we consider different subsets. This means that the properties we find are not tied to individual neurons, but are general features of the ensemble. Since collective synchrony is a lower-dimensional statistic and in principle permits exchange symmetries between neurons, this is not necessarily surprising. On the other hand, we note that these collective properties are robust to heterogeneity in network connectivity, bias, and interactions that would tend to break such symmetries.

\section*{Discussion}
Examples of how sensory and behavioral information is encoded in neural activity suggest that, while information is sometimes localized to a few neurons, it is at other times distributed amongst many. This may result from information flow between collective to individual components \cite{danielsDualCoding2017} and because the precise scale at which information is processed may vary with time, function, and organism. We develop a perturbative approach that is sensitive to the potential variety of involved scales. 
Using an information-geometric perspective, we identify statistical aspects of the distribution of neural behavior that are essential for preserving collective activity.
Importantly, we do not have to assume that collective properties will be sensitive to individual neurons or certain combinations but can discover the appropriate ones in a principled way. This becomes feasible to explore comprehensively because the response of the system depends on only pairs of perturbations when they are sufficiently small, a property of analyticity of the mapping from activity to collective state.

As an {\it in silico} realization of the protocol, we use perturbations that mimic internal neural coordination and calculate their impact on collective synchrony. In a minimal model, we find that dominant perturbative modes do not tend to be distributed amongst many neurons but are localized into pivotal ones. The concentration of sensitivity in a few neurons is analogous to the presence of driver neurons, modulation of which drives the system from one configuration to another \cite{mitraWmatrixGeometry1969,royRelationFIM2009,liuConvergenceFundamental2006}. Interestingly, we find that pivotal neuron identities fluctuate across subsets and Monte Carlo random samples that introduce finite sample errors into the entries of the FIM. Instead of specific pivotal neurons, it is the collective properties of localized eigenvectors and scaling in sensitivity that are preserved.

These collective properties require neural heterogeneity. Since identical neurons would imply uniform eigenvectors, variation in each neuron's local interaction network and bias is responsible for the emergence of pivotal neurons \cite{leeSensitivityCollective2020}. By shuffling the inferred interaction network and the data, we verify that these features seem to be a result of the distribution and magnitude of interactions but not the specific topology. In this sense, the neurons do not need to be labeled differently as if given some predetermined role, but the collection of local interactions and biases is responsible for the statistical properties of the FIM, echoing findings in theoretical models of neural networks \cite{rajanEigenvalueSpectra2006}. Though we do not address this question here, it would be interesting to know how such properties emerge and if such ensemble properties are encoded genetically, arise spontaneously, or generalize to mobile worms.

One of major questions that we approach in our framework is how experimental perturbations should be represented in the model. While the focus is often on model parameters as representations of physical dials that are experimentally accessible, direct correspondence is not obvious for a statistical model inferred from data. As an example, when couplings inferred by a pairwise maxent model were directly compared with physical contact between protein residues, the model recovered only a subset of real interactions even while recovering the statistical ensemble \cite{morcosDirectcouplingAnalysis2011,weigtIdentificationDirect2009}. Taking the pairwise maxent model, it is clear exactly what increasing a coupling does to the energy function, enhancing the tendency of a pair of neurons to coincide, but the actual result is nontrivial modification of the entire distribution. For the experimentalist, it seems more natural to consider the problem from the perspective of the observable statistics that can be perturbed in a controlled and precise manner. In the case of Boltzmann-type models, the relationship between observables and model parameters can be made exact and results in a simple form for small perturbations. Thus, our formulation is a theoretically and experimentally tractable framework for predicting the effects of perturbations.

To test these ideas requires innovations on top of previous experiments (Section~\ref{si sec:experiment}). A central feature of our thought experiment is the small perturbation. If the strength of the intervention ranges from perturbative to discrete, the perturbative limit lies close to no intervention at all and discrete interventions correspond to silencing or ablation. Discrete interventions may be appropriate considering the logical structure of a circuit but become combinatorially expensive to explore fully \cite{oliveriGlobalRegulatory2008,liEncodingRegulatory2014}. While avoiding these issues, small perturbations require a sufficient number of measurements for the perturbed parameters to be measurable. By using the fact that the Fisher information is proportional to the number of independent samples and the Cram\'{e}r-Rao bound, a lower bound on the number of experimental observations $T$ required to distinguish a change in the mode for a perturbation strength $\epsilon$ for Fisher information $F$ is on the order of
\begin{align}
	T \sim (\epsilon^2 F)^{-1}.
\end{align}
Taking a relatively large perturbation of $\epsilon = 10^{-2}$, we have $T\sim10$ independent samples for an eigenvalue $\lambda \sim 10^3$, the order of magnitude of the largest mode \cite{fim_footnote}. Linear growth in the number of required samples, however, restricts us from measuring insensitive modes in a reasonable amount of time (the experiments analyzed here last 8 minutes and collect roughly 80 to 120 independent samples \cite{chenSearchingCollective2019}). Though mapping the full local information geometry of $N=50$ neurons would be difficult because the number of pairwise perturbations exceeds $\sim10^6$ (accounting two distinct pairs of matcher and target neurons), a similar experiment with about $N\sim10$ neurons seems reasonable when coupled with techniques for incomplete matrix estimation \cite{candesExactMatrix2009}. The feasibility of such an experiment stems from our argument that perturbative experiments harness analyticity to vastly simplify the range of possible perturbations. We exploit this property to extract a basis for the local information geometry including multi-component perturbations. One goal of such experimental intervention is then not to be more precise but sufficiently varied to span the local basis, an idea that can be generalized to other experimental systems besides the \celegans\ model we consider. 

Until now, we have assumed that the stochastic encoding relating neural to collective activity does not change, but this is not necessarily the case. To consider this, we characterize the encoding with the fundamental notion of channel capacity, the maximum rate at which information can conveyed by such a mapping given by the mutual information $I[Y;Z] = \sum_{Y,Z} P(Y,Z) \log \left(P(Y,Z)/P(Y)P(Z)\right)$ \cite{shannonMathematicalTheory1948}. A targeted perturbation to the set of states $Y$ corresponds to a new distribution $\tilde P(Y) = P(Y)[1+ a_r v_r(Y)]$ giving arbitrary weight $a_r$ to perturbation vector $v_r$ of index $r$ \cite{boltzmann_note}. The perturbation vector $v_r$ may refer to a projection from a single-neuron perturbation or eigenvector considered above. Generally, such a perturbation may change not only the distribution of neural statistics $P(Y)$ but also the mapping $\tilde{P}(Z|Y) = P(Z|Y)[1+\Delta_{Z|Y}]$, which corresponds to an adaptive code that changes in response to incoming statistics. Under small perturbations $a_r\ll 1$ and small adaptation $\Delta_{Z|Y}$, the change in the mutual information decomposes into
\begin{align}
	\tilde{I}[Y;Z] - I[Y;Z] &= \underbrace{a_r\sum_{Y} P(Y) v_r(Y) \sum_Z d(Z|Y)}_{\rm direct} + \underbrace{\sum_{Y}P(Y) \sum_Z \Delta_{Z|Y}d(Z|Y)}_{\rm adaptive}, \label{eq:delta I}
\end{align}
where we have defined $d(Z|Y)\equiv P(Z|Y)\log [P(Z|Y)/P(Z)]$ such that $\sum_Z d(Z|Y)$ is the information gained about $Z$ from observing $Y$. Eq~\ref{eq:delta I} contrasts the ``direct'' result of a perturbation in contrast with an ``adaptive'' response by the system. Whereas the former term decomposes into the product of the local information geometry of $P(Y)$ and the information content of the mapping $Y\rightarrow Z$, the latter depends on system response that modifies the stochastic mapping itself, such as rate limiting or rescaled response \cite{brennerAdaptiveRescaling2000}. Deviations from the predictions in our perturbative thought experiment might represent the effects of such complications from the properties of the stochastic encoding.

The information geometry of scientific theories more generally suggests that they show sparse structure, where a few parameter dimensions strongly change the qualitative characteristics of phenomena and most dimensions are unimportant. This is a result of the logarithmically spaced eigenvalues of the FIM, or ``sloppiness'' \cite{transtrumGeometryNonlinear2011,transtrumWhyAre2010,transtrumPerspectiveSloppiness2015,machtaParameterSpace2013}. Whether by nature or by design, such quantitative reduction vastly simplifies the level of detail required for approximate theories, allowing for accurate prediction even with large uncertainty in most parameters \cite{danielsParameterEstimation2018,transtrumModelReduction2014}. We find here a variation on this idea in the statistics of neural activity in \celegans. Neural activity seems non-sloppy, with largest eigenvalues decaying slower like Zipf's law in contrast with regular logarithmic spacing when we consider experimentally realistic perturbations. Thus, sensitivity, while concentrated into a few dominant modes, has non-negligible smaller modes that decay in a self-similar way. Might this be a feature of biological neural networks to reduce the dimensionality of control parameters (if interestingly not as strongly as the sloppy case) or to nest varying levels of control? The state-of-the-art today with single neuron measurement in \celegans\ and optical genetics is approaching a point where experiments to test this hypothesis may become feasible.

\section*{Acknowledgments}
We would like to thank the UNM Center for Advanced Research Computing, supported in part by the National Science Foundation, for providing high performance computing resources used in this work. We thank Matthew Fricke for invaluable assistance with these resources. We thank the Santa Fe Institute for providing computational resources used for this work. The computational results presented have been achieved in part using the Vienna Scientific Cluster (VSC). E.D.L.~acknowledges funding from the Omega Miller Program, NSF grant PHY1838420, and Medical University of Vienna. X.C.~acknowledges Francesco Randi for insightful discussion. B.C.D.~was supported by the ASU–SFI Center for Biosocial Complex Systems.

\nolinenumbers

\clearpage
\appendix
\renewcommand\thefigure{\thesection.\arabic{figure}}

\section{Numerical solutions to inverse maxent problem}\label{si sec:solution}
In principle, the maxent formulation presents a unique mapping from statistical correlations to model parameters such that there is no parameter fitting in the usual sense. 
The problem of determining the fields $h_{i,k}$ and couplings $J_{ij}$ that closely match mean activity and pairwise correlations is known as the inverse maxent problem \cite{nguyenInverseStatistical2017,leeConvenientInterface2019}. In practice, however, limitations to numerical precision and finite-sample noise mean that fitting the parameters for even moderately sized systems is not without ambiguity. With this ambiguity in mind, we present the different approaches we use to solve the inverse problem for the pairwise maxent and independent models. 

\begin{figure}\centering
	\includegraphics[width=\linewidth]{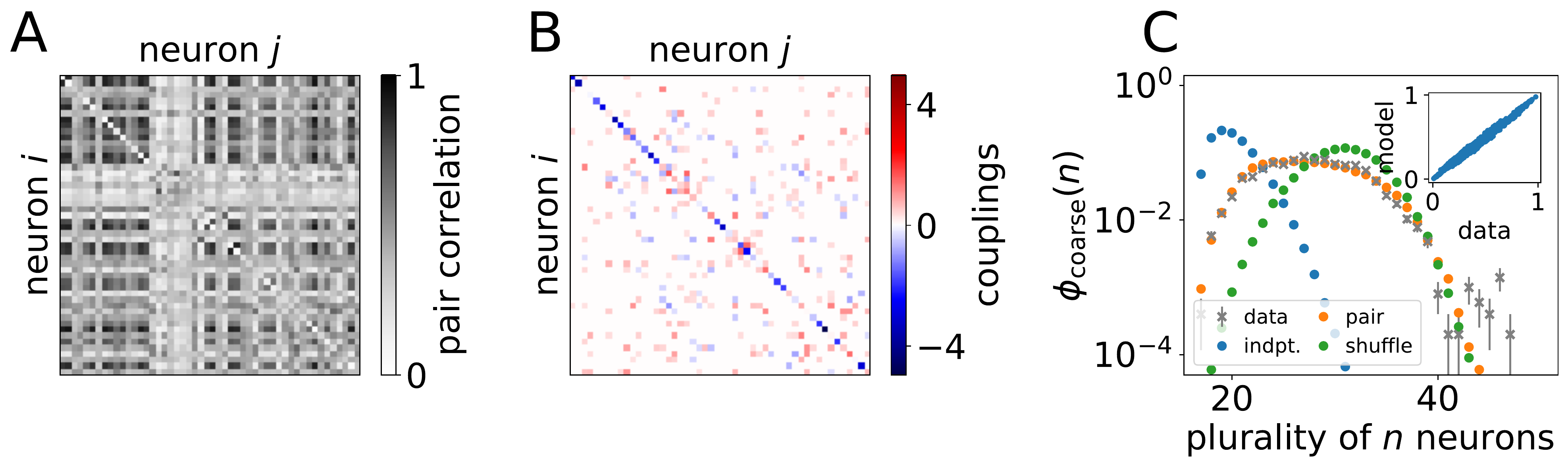}
	\caption{Pairwise maxent model of \celegans\ anterior neural activity for experiment 170419 from reference \cite{scholzPredictingNatural2018}. (A) Pairwise correlations between subset of $N=50$ neurons. Average individual neuron $r_{k=-1}(s_i)$ shown along diagonal. (B) Inferred biases $h_{i,k=-1}$ along diagonal and interaction matrix of couplings $J_{ij}$ off the diagonal. (C) Collective synchrony, probability that a plurality of $n$ neurons coincide $\phi_{\rm coarse}(n)$. Inset compares pairwise correlations for data, pairwise maxent model, independent model, and one example of shuffled couplings. See reference \cite{chenSearchingCollective2019} for more details about the model. See Figure~\ref{si gr:maxent second} for experiment 154139. Error bars show one standard deviation over bootstrapped samples.}\label{si gr:maxent}
\end{figure}

\begin{figure}\centering
	\includegraphics[width=\linewidth]{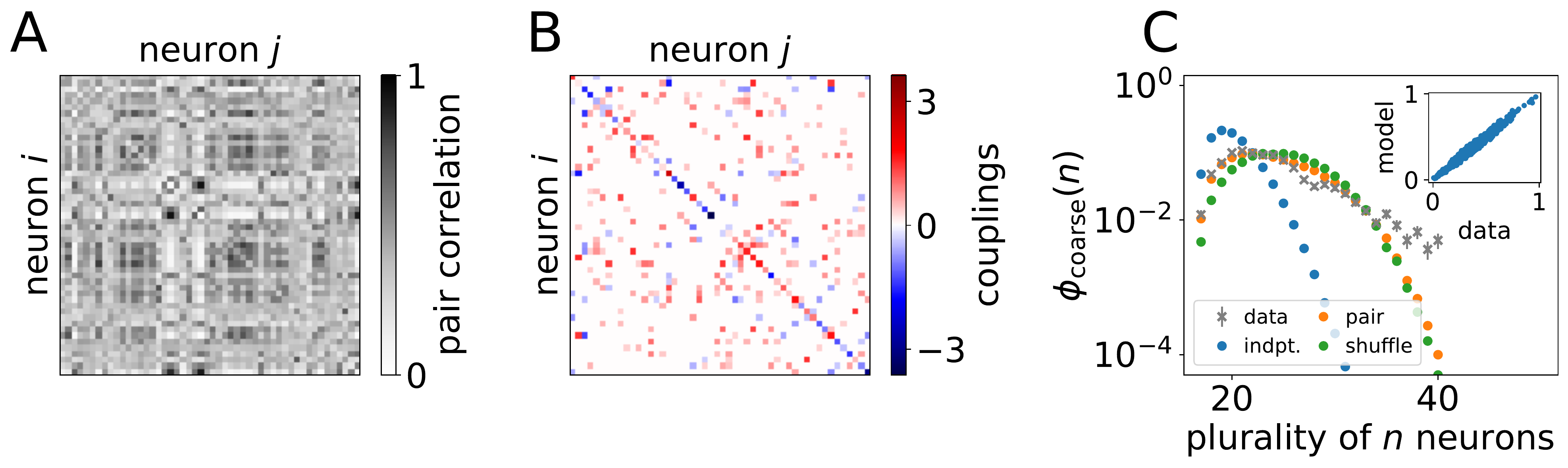}
	\caption{Maxent model fit overview for experiment 154139. See Figure~\ref{si gr:maxent} for experiment 170419 and explanation of panels.}\label{si gr:maxent second}
\end{figure}

The first method, which we focus on in the main text, allows us to identify a sparse interaction network between neurons mirroring the sparse structural connectivity of the neural connectome \cite{chenSearchingCollective2019}. The parameters of the maxent model are initialized at zero and are updated individually as to maximally increase of the likelihood of the data at each step \cite{dudikPerformanceGaurantees2004, broderickFasterSolutions2007}. This learning procedure is stopped once the model reproduces the constrained observables within experimental variability, estimated by bootstrapping random halves of the data. If experimental variability is relatively large, this training procedure can return a sparse interaction matrix $J_{ij}$. We show the results of such a procedure in Figures~\ref{si gr:maxent} and \ref{si gr:maxent second}.

The second method we use for comparison relies on the Monte Carlo Histogram (MCH) approach \cite{broderickFasterSolutions2007}, which is an approximate gradient descent algorithm. At each step, the parameters are simultaneously updated by an amount that is proportional to the difference between the corresponding observable calculated from the model and the data. We obtain a solution that is within a norm error threshold
\begin{align}
	\sqrt{\sum_{i=1}^N (\br{\si}_{\rm data} - \br{\si})^2 + \sum_{i<j}^N(\br{\si\sj}_{\rm data} - \br{\si\sj})^2}<1/2,
\end{align}
where the cutoff is arbitrarily set to obtain a relatively fast and close fit to the statistics of the data. Unlike the first method, MCH returns a dense network of connections since all couplings are updated at every iteration til convergence. While we find that the resulting model aligns well with the features of the data, including collective synchrony, this procedure does not make realistic assumptions about the topological structure of the underlying physical network. For these solutions, we again find a signal for distinguishing pivotal neurons in the uniformity of columns and rows, but it is not as consistent as evident in the column and row uniformities plotted in Figure~\ref{si gr:uniformity mch}.

We also consider an independent neuron model. When solving the corresponding maxent model, we penalize large fields by maximizing the log-likelihood $\log\mathcal{L}_i$ for each spin $i$ along with a penalty such that the total cost function $\mathcal C_i$ is defined as
\begin{align}
	\mathcal{C}_i &\equiv -\log\mathcal{L}_i + \frac{1}{\sigma}\sum_{k=-1}^1 h_{i,k}^2\label{si eq:indpt cost}
	\intertext{because large fields make it especially costly to calculate the FIM accurately. Indeed, a sparse cost function, one where the cost scales with the absolute value of the fields, does not sufficiently penalize large fields, rendering it infeasible for our current implementation of the FIM calculation. With the constraint given in Eq~\ref{si eq:indpt cost}, the goal is then to find the fields such that}
	h_{i,k}^* &= \min_{h_{i,k}} \mathcal{C}_i.
\end{align}
To determine the weight $\sigma$, we compute the cost function $\mathcal{C}_i$ in Eq~\ref{si eq:indpt cost} for a range of $\sigma$. At small $\sigma$, the quadratic penalty dominates and the function approaches a large constant. At large $\sigma$, we recover the original maximum likelihood problem. We set $\sigma=\sigma^*$ as is determined by the midpoint between these two extremes for the cost function averaged over all spins $i$. Typically, this is in the interval $\sigma^*\in[2,10]$. These steps return an independent model of neurons, where the biases of the most extreme neurons have been tempered.

\section{Calculation of Fisher information matrix (FIM)}
To calculate the entries of the FIM, we rely on a Monte Carlo Markov Chain (MCMC) sample of the distribution of neural states $p(s)$, which we then coarse grain to approximate the distribution over collective synchrony $\phi$. With the distribution, we then calculate its Kullback-Leibler divergence under perturbation, which is straightforward to do by altering the statistical correlation functions calculated on the MCMC sample according to Eq~\ref{eq:replacement rule}. Relating this change to a change in the fields $\{h_i\}$ and couplings $\{J_{ij}\}$ is a linear matrix problem in the perturbative limit. With the corresponding change in the parameters, we calculate the change in energy $\Delta E(\{s_i\})$ for each configuration $s$ sampled, the coarse graining of which maps onto the corresponding effective energy of the system. This algorithm is specified in more detail in the supplementary information of reference \cite{leeSensitivityCollective2020}.

\begin{figure}[tbp]
	\centering
	\includegraphics[width=.6\linewidth]{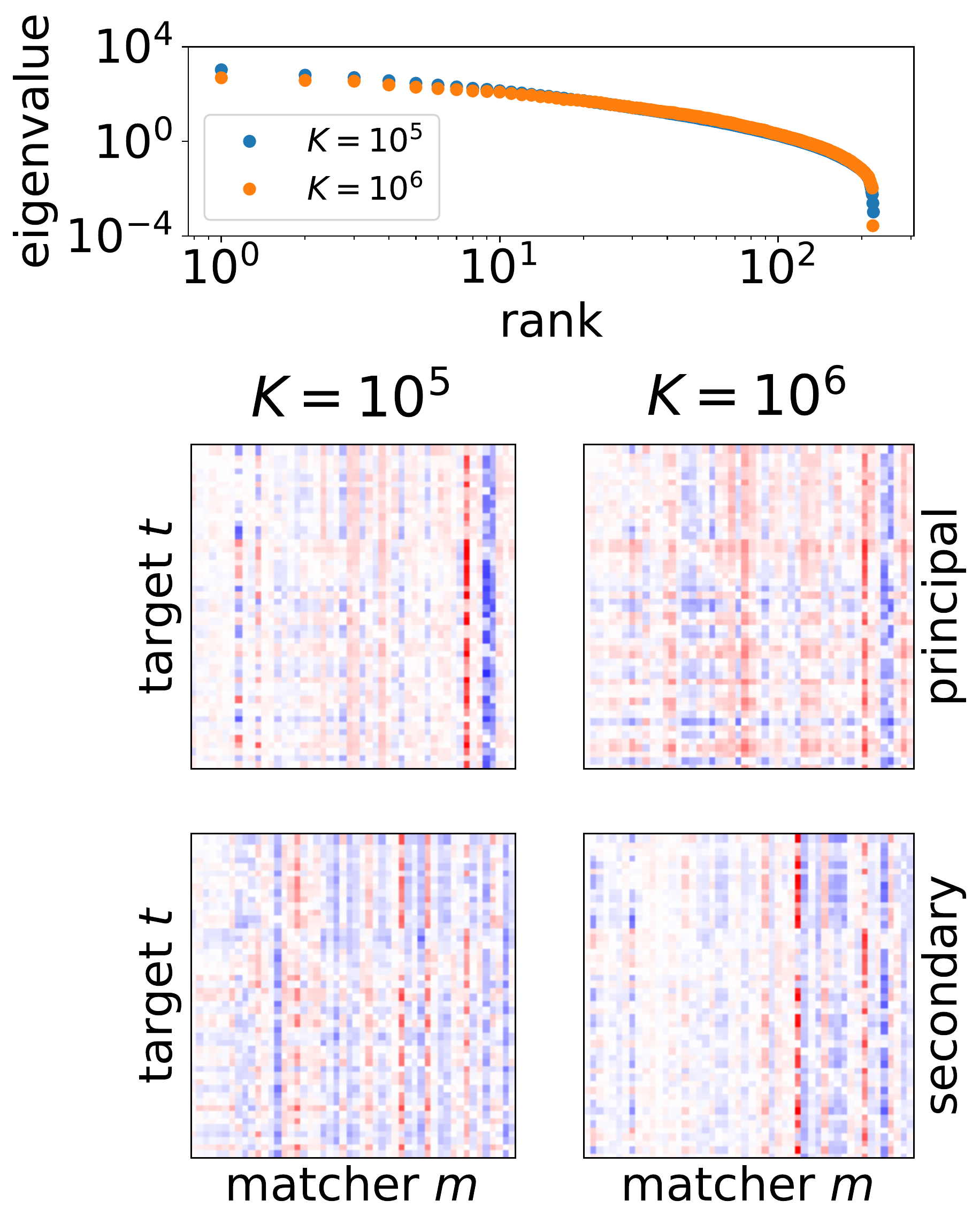}
	\caption{Comparison of (top) eigenvalue spectrum and (bottom) eigenmatrices calculated from Monte Carlo samples of sizes $K=10^5$ and $K=10^6$. Eigenvalue spectrum for $K=10^5$ is an average calculated over 10 different Monte Carlo samples. Top two eigenvectors are taken from a single sample as an example, and the two columns share similar, if not identical, features.}\label{si gr:sample size}
\end{figure}

Overall, the calculation of the FIM is an expensive computational task where we must obtain each entry of the $(N^2-N)\times(N^2-N)$ FIM which are averages over $K$ MCMC samples for each of $|\phi|$ coarse-grained statistics. To parallelize such a calculation, we combine custom code with the Metropolis algorithm implemented in the ConIII Python package \cite{leeConvenientInterface2019}. We relied on a number of computational resources including local workstations and computing clusters at the Santa Fe Institute, University of New Mexico, and Vienna Scientific Computing (VSC).

\begin{figure}\centering
	\includegraphics[width=.9\linewidth]{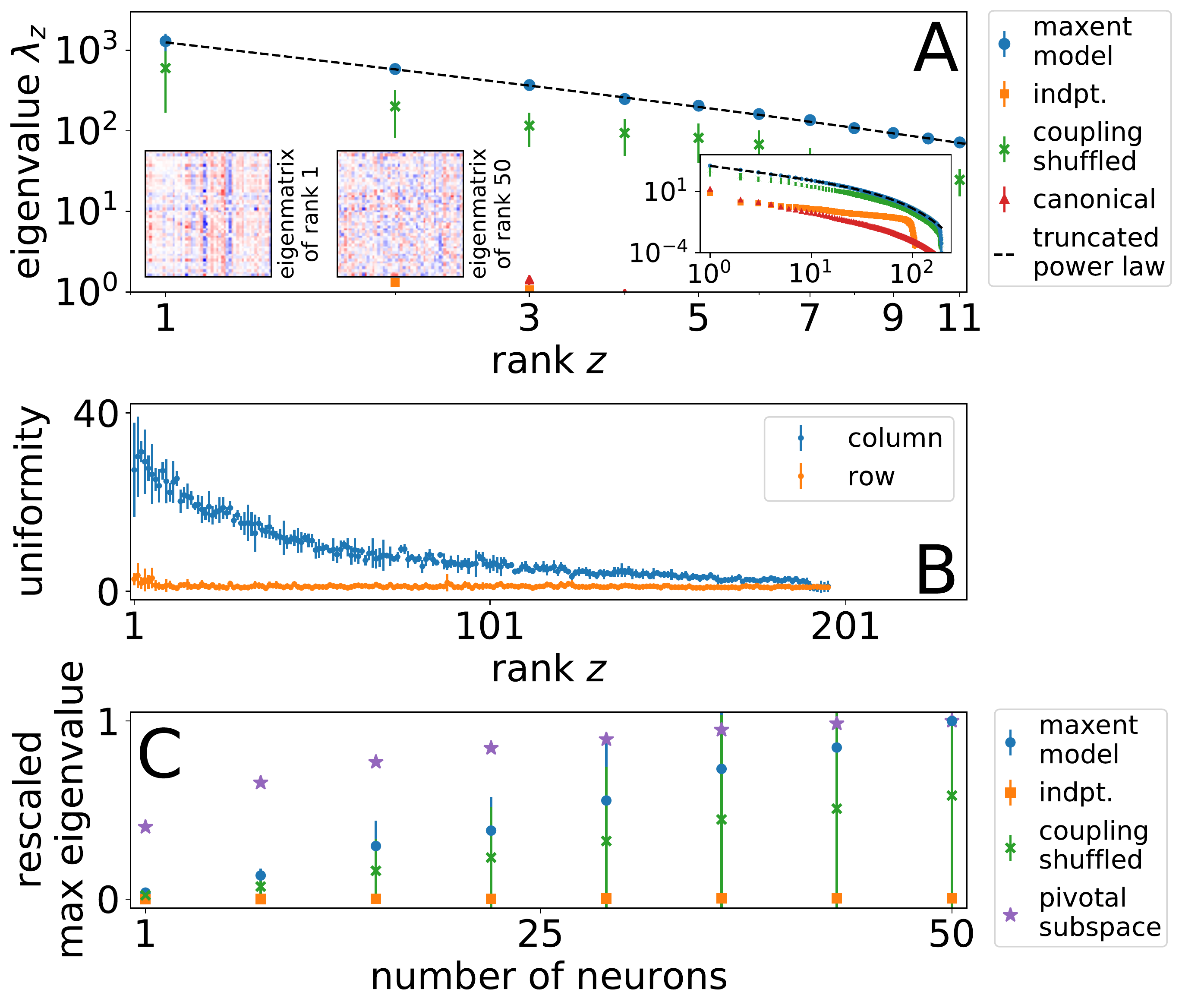}
	\caption{FIM overview for worm experiment 154139. In main text, we show corresponding plots for experiment 170419 in Figure~\ref{gr:eigenspectrum}.}\label{si gr:eigenspectrum}
\end{figure}

In the main text, we show results from MCMC samples of size $K=10^5$ and compare a few cases with a larger samples of size $K=10^6$ to verify that we can obtain a good approximation of FIM properties with the smaller sample having fixed $\epsilon=10^{-4}$.\footnote{At the very least, perturbation magnitude should be smaller than the inverse of the MCMC sample size.} We verify that our estimates of the FIM entries converge within a relative tolerance of $0.1\%$ when compared with a larger perturbation of $\epsilon=2\times10^{-4}$. The example of the eigenvalue spectra in Figures~\ref{gr:eigenspectrum} and \ref{si gr:eigenspectrum} involve an average over 10 samples of size $K=10^5$ for the pairwise maxent model the shuffled null model. See Figure~\ref{si gr:sample size} for a comparison of spectral properties of the FIM between the two sample sizes considered.

\section{Analyzing eigenmatrices}\label{si sec:fim analysis}
\begin{figure}\centering
	\includegraphics[width=\linewidth]{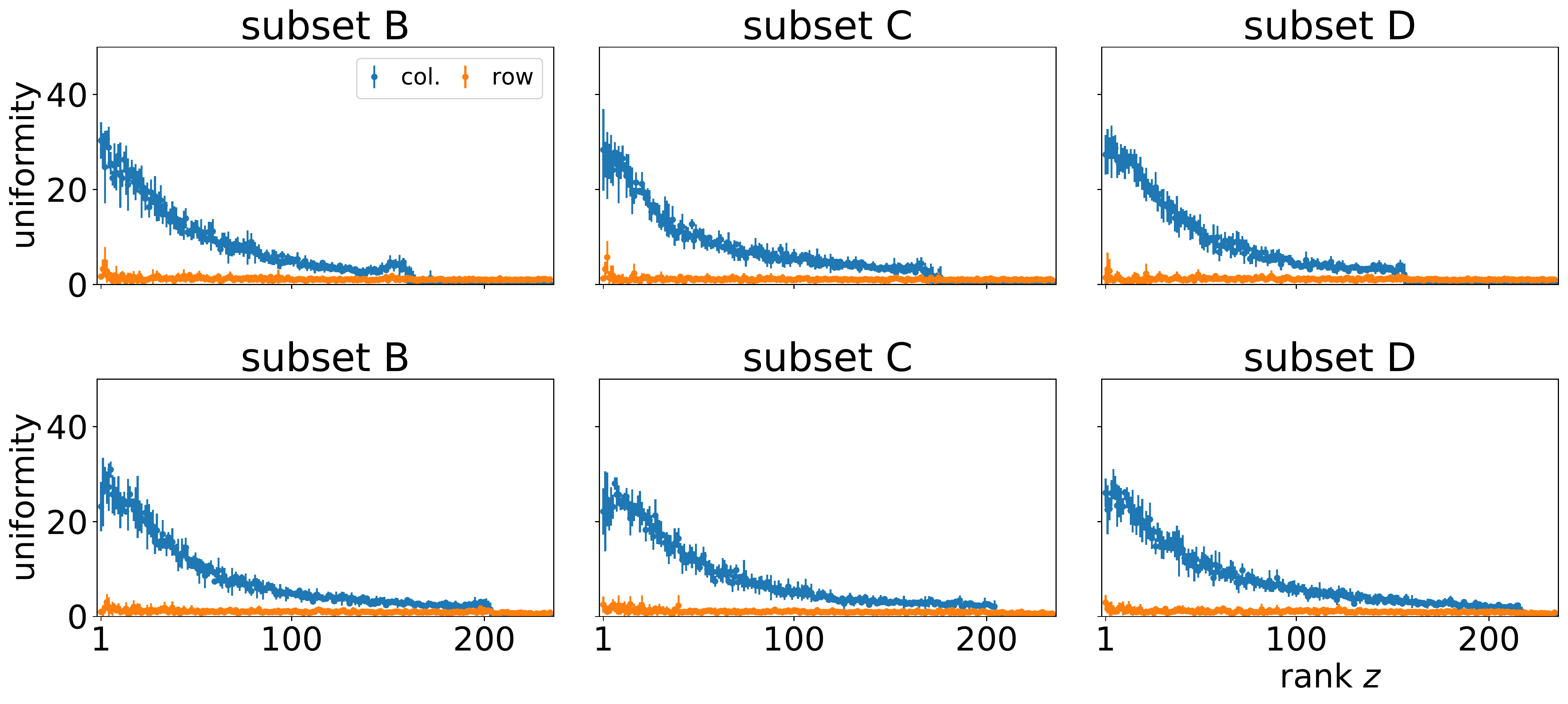}
	\caption{Uniformity of eigenmatrices for different neural populations of $N=50$ neurons. We show averages over Monte Carlo samples. Error bars show one standard deviation. (top row) Subsets from experiment 154139. (bottom row) Subsets from experiment 170419.}\label{si gr:uniformity}
\end{figure}

\begin{figure}\centering
	\includegraphics[width=\linewidth]{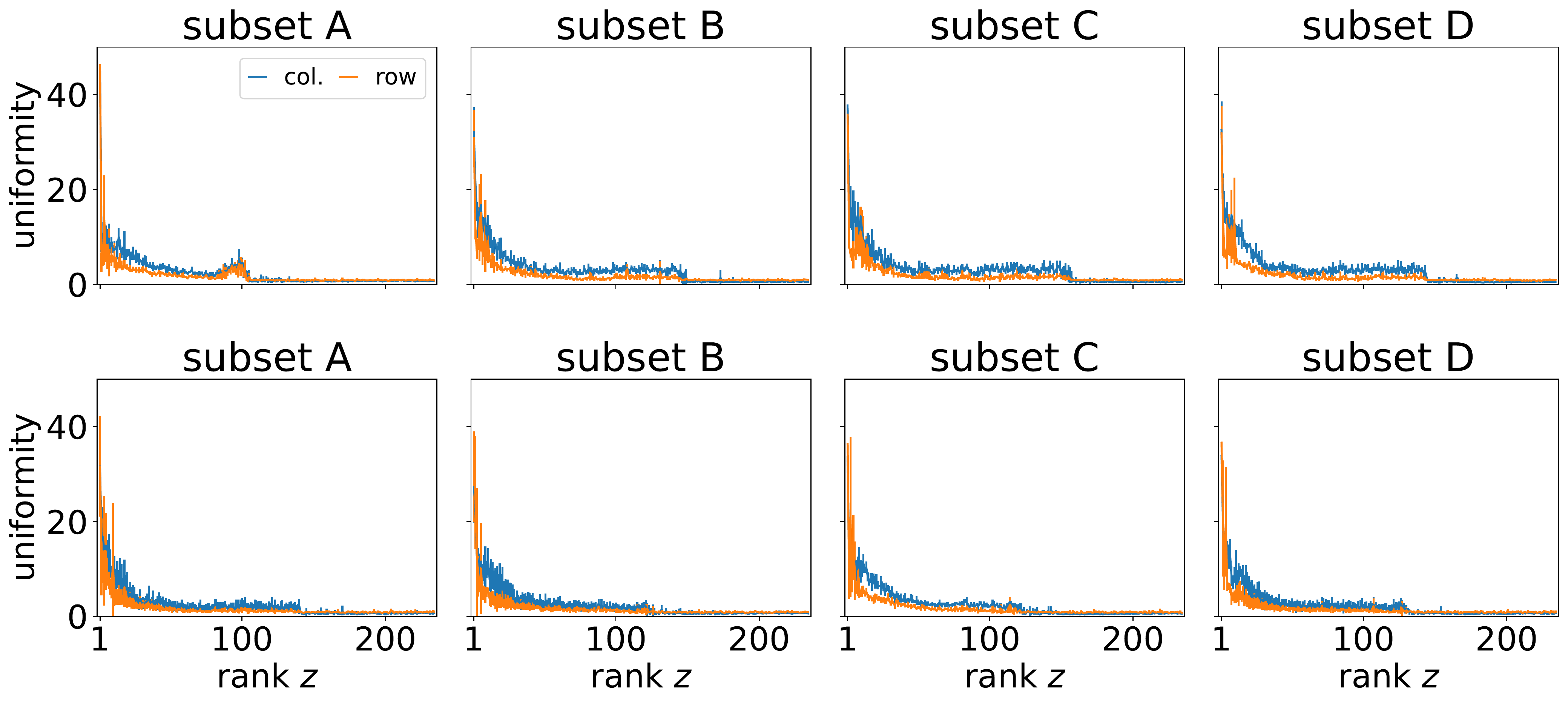}
	\caption{Uniformity of the eigenmatrices for independent model. There is no dramatic separation between column and row uniformity. See Figure~\ref{si gr:uniformity} for details.}\label{si gr:uniformity indpt}
\end{figure}

\begin{figure}\centering
	\includegraphics[width=\linewidth]{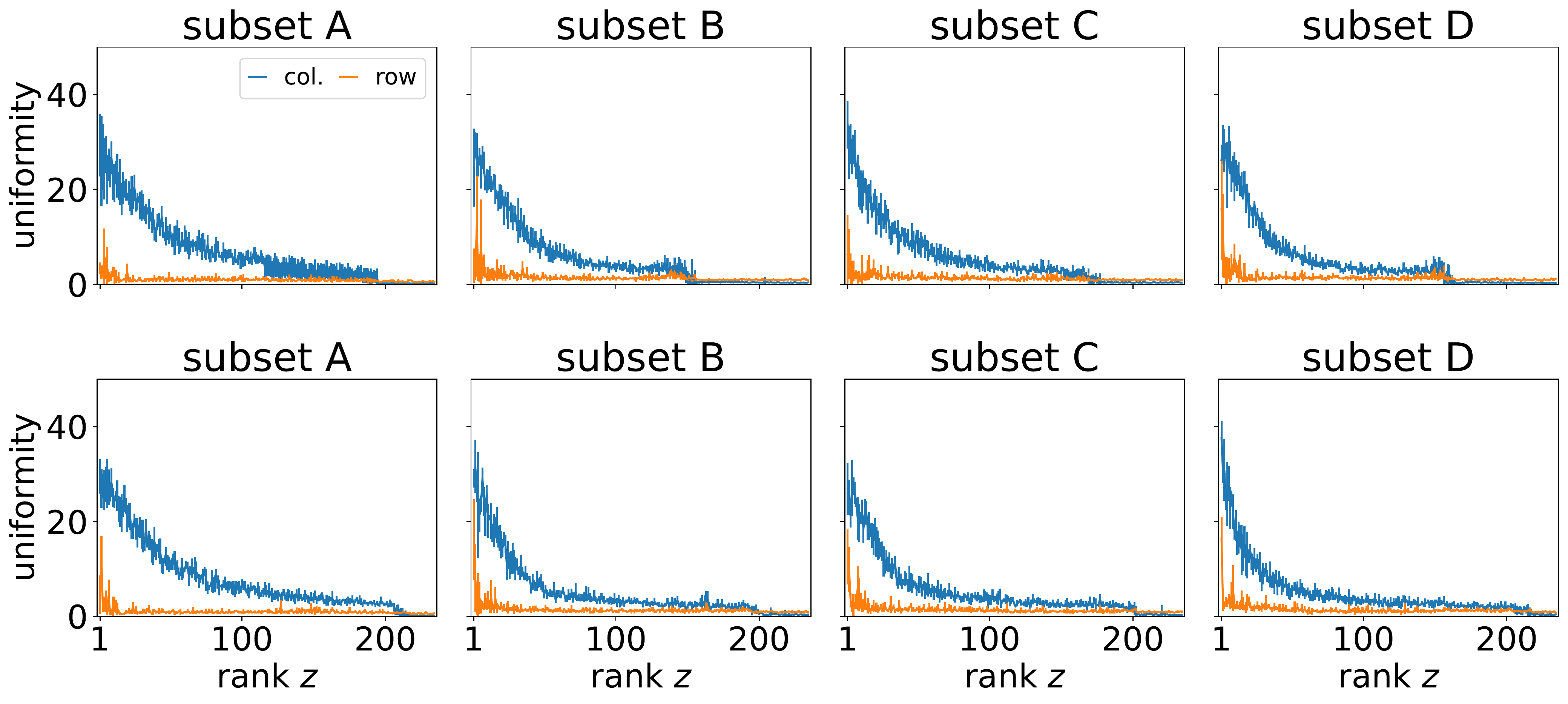}
	\caption{Uniformity of eigenmatrices after shuffling couplings. Permutation mostly preserves separated column and row uniformities. See Figure~\ref{si gr:uniformity} for details.}\label{si gr:uniformity shuffle}
\end{figure}

\begin{figure}\centering
	\includegraphics[width=\linewidth]{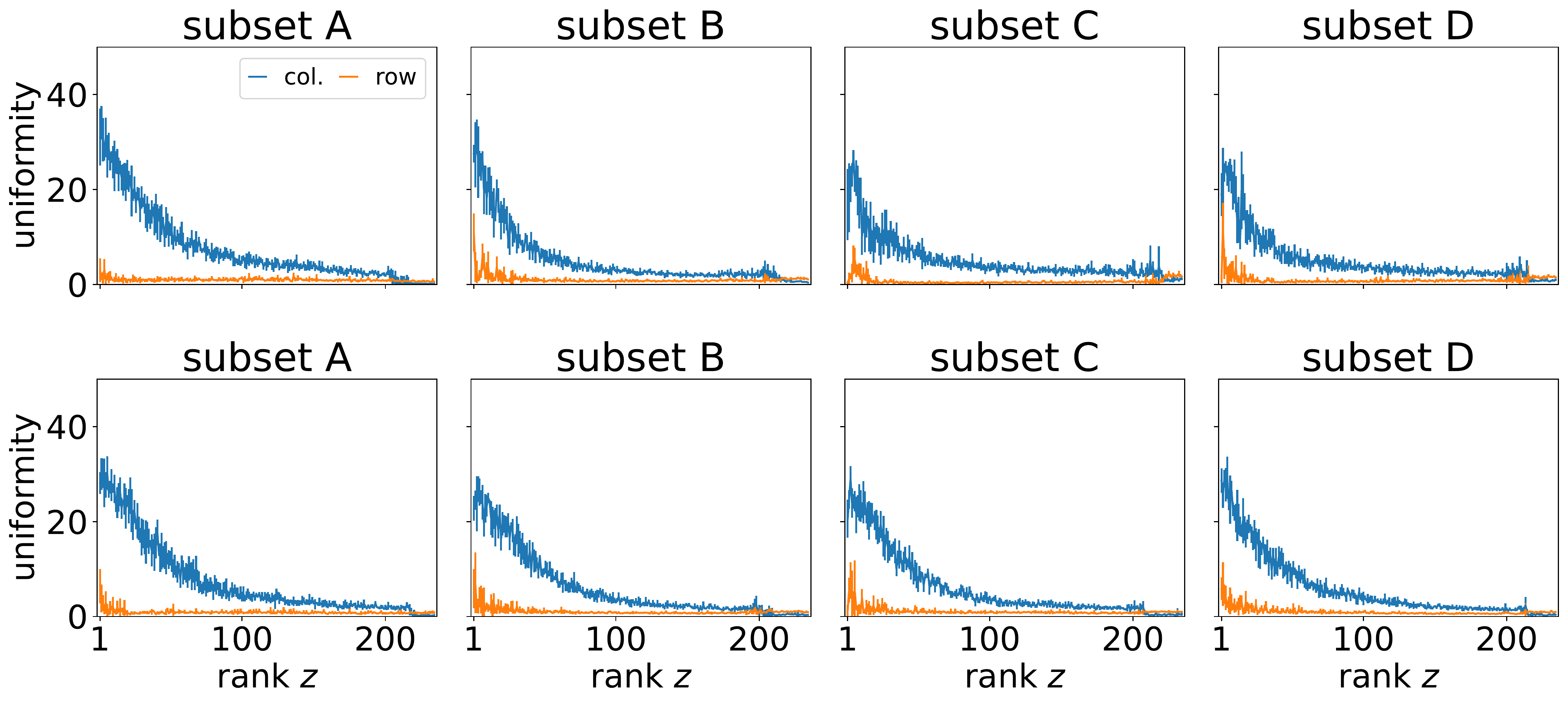}
	\caption{Uniformity of eigenmatrices for MCH solutions. See Figure~\ref{si gr:uniformity} for details.}\label{si gr:uniformity mch}
\end{figure}

We compare vertical striations in comparison withrows by measuring comparing the sum of row and column uniformities as defined in Eq~\ref{eq:uniformity}. In Figures~\ref{si gr:uniformity}, \ref{si gr:uniformity indpt},\ref{si gr:uniformity shuffle}, and \ref{si gr:uniformity mch}, we show uniformity for 3 additional neuron subsamples for the main approach discussed in the text, the independent model, the shuffled null model, and the MCH solutions, respectively. For pairwise perturbations, we find matrices strongly biased to large column sum norms compared to rows. Given the nature of the pairwise perturbations that we consider, that means that perturbations localized to a particular neuron (in contrast with perturbations that would impact each of the neighbors in turn) would have a dominant impact on the collective outcomes. This is a feature of localized control because it means that turning off all local synaptic connections, or turning them up, for a particular neuron is effective. 

\begin{figure}\centering
	\includegraphics[width=.65\linewidth]{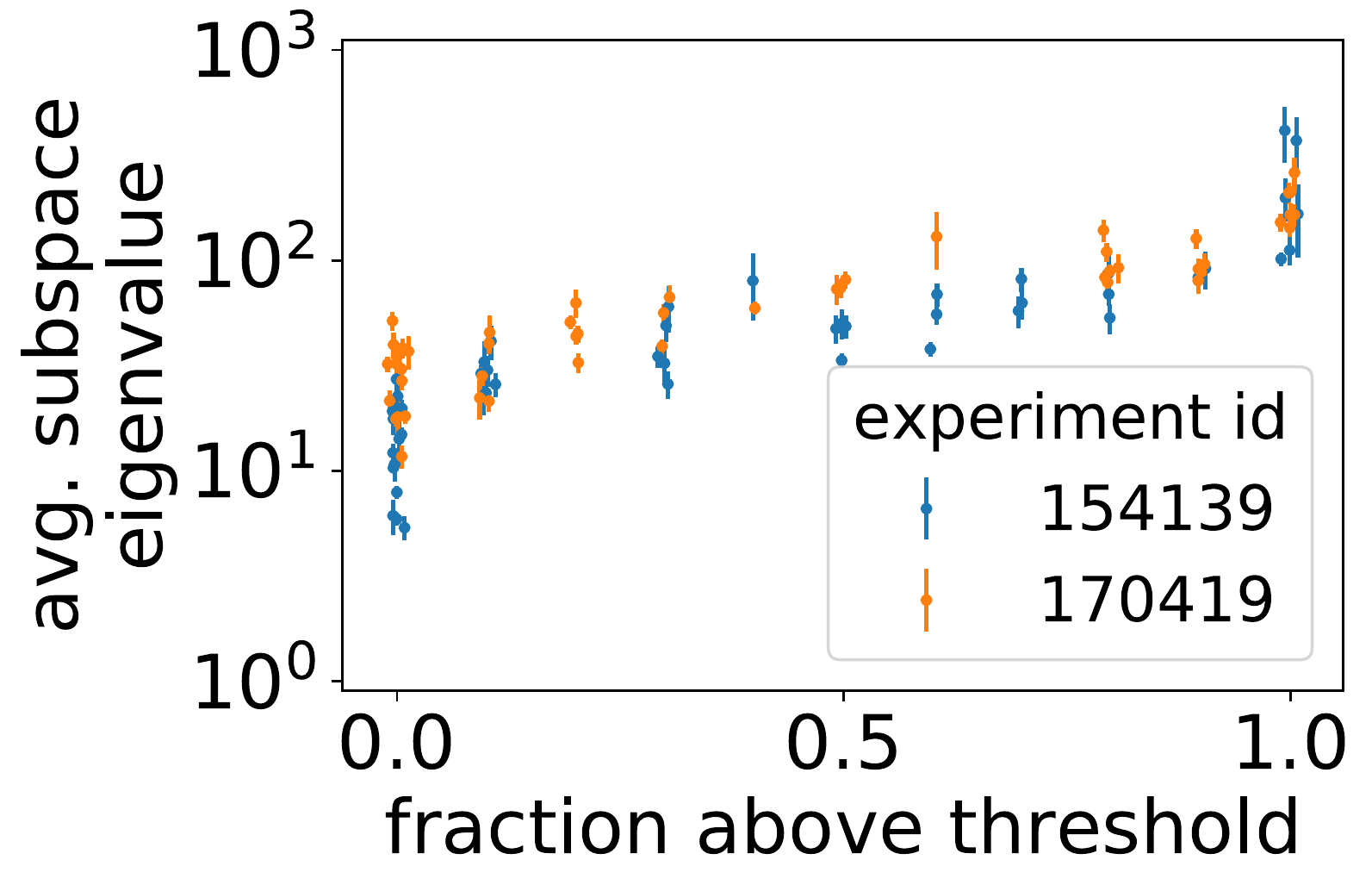}
	\caption{Principal subspace eigenvalues vs.~fraction above column uniformity threshold of 99-percentile (compare with Figure~\ref{gr:piv neurons}). Subspace corresponds to diagonal blocks of FIM for fixed matcher index $m$, which is used in Figure~\ref{gr:eigenspectrum}C to order neurons. Average taken over ten MC samples with standard error of the mean shown. This means that neurons often above threshold tend to have large collective sensitivity, linking these two different measure of pivotal strength. Points are randomly offset on x-axis for visibility.}\label{si gr:piv measures}
\end{figure}

As a more direct measure of this, we can analyze the subspace of the FIM corresponding to perturbation of each neuron at a time, fixing $m$ and iterating over all $t$. As we show in Figure~\ref{si gr:piv measures}, the principal eigenvalues extracted from single neuron perturbations are correlated with the fraction of MC samples in which the same neuron has unusually strong column uniformity. Thus, two different measures of pivotal neurons, one based on single neuron subspaces and another based on the structure of the entire system, align and show that the strong collective tendencies that we find in the data do not prevent individual neurons from playing important roles.

\begin{figure}\centering
	\includegraphics[width=.75\linewidth]{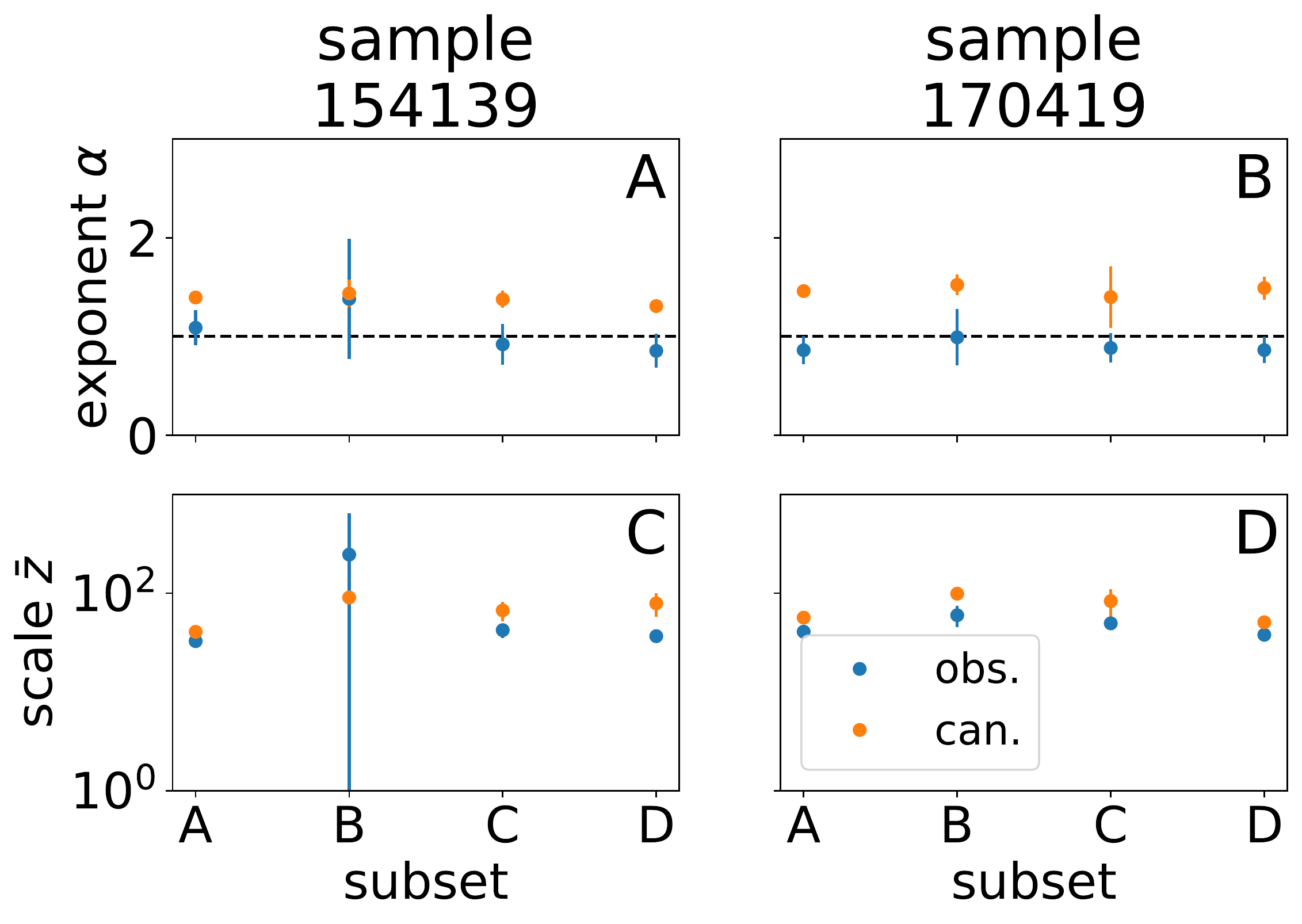}
	\caption{(A, B) Power law exponent from fitting FIM eigenvalue spectra comparing observable and canonical perturbations. (C, D) Scale of exponential tail $\bar{z}$.}\label{si gr:exponent fits}
\end{figure}

To determine the scaling of the rank-ordered eigenvalue spectrum, we fit the function detailed in the main text and reproduced here
\begin{align}
	\lambda(z) &= C z^{-\alpha} e^{-z/\bar{z}}\label{si eq:eig fit}
\end{align}
for eigenvalue $\lambda(z)$ with rank $z$ by performing least-squares minimization on the logarithmic differences. However, there no {\it a priori} guarantee that the spectrum of eigenvalues is full rank and the tail of the spectrum may be rife with numerical precision errors, and so we allow for the possibility of a hard cutoff that should be applied. As a heuristic, we apply a hard cutoff when the logarithmic slope falls below $-3$ and do not fit any points for rank above the cutoff. While we anticipate that numerical precision makes it difficult to estimate the smallest eigenvalues and thus the exact value of the cutoff, we take the sharp tail to be an effective cutoff for when the eigenvalues fall below a threshold of $\lambda<10^{-7}$, minuscule in comparison with the typical principle eigenvalue in the range of $10^2$ to $10^4$. As is visible in our fit in Figure~\ref{gr:eigenspectrum}, Eq~\ref{si eq:eig fit} with a hard cutoff hews very well to the averaged FIM spectrum. In contrast, a simple exponential decay, having set $\alpha=0$, cannot capture the scaling form that we observe. We show an overview of the truncated power law fit exponents and exponential tails in Figure~\ref{si gr:exponent fits}.

We compare our findings for the pairwise maxent model with null models including an independent model for neural activity and one where the couplings are randomly assigned to a pair called ``coupling shuffled.'' While the independent model gives qualitatively different results, we find that shuffling the couplings amongst pairs preserves the qualitative features of the FIM.

\section{Classes of perturbations}
We distinguish between two principal classes of perturbations in the main text denoted ``observable'' and ``natural'' or ``canonical,'' taking the nomenclature used in Amari's well-known textbook on information geometry \cite{amariInformationGeometry2016}. Essentially, these distinguish between perturbations defined in the space of correlations compared with perturbations defined in the space of parameters. Since perturbations in either representation can be transformed to one other by a linear operation, there is no {\it a priori} reason to prefer one basis over another. Theoretical treatments tend to consider the canonical picture because of the underlying intuition that they correspond to physical forces.

\begin{figure}\centering
	\includegraphics[width=.75\linewidth]{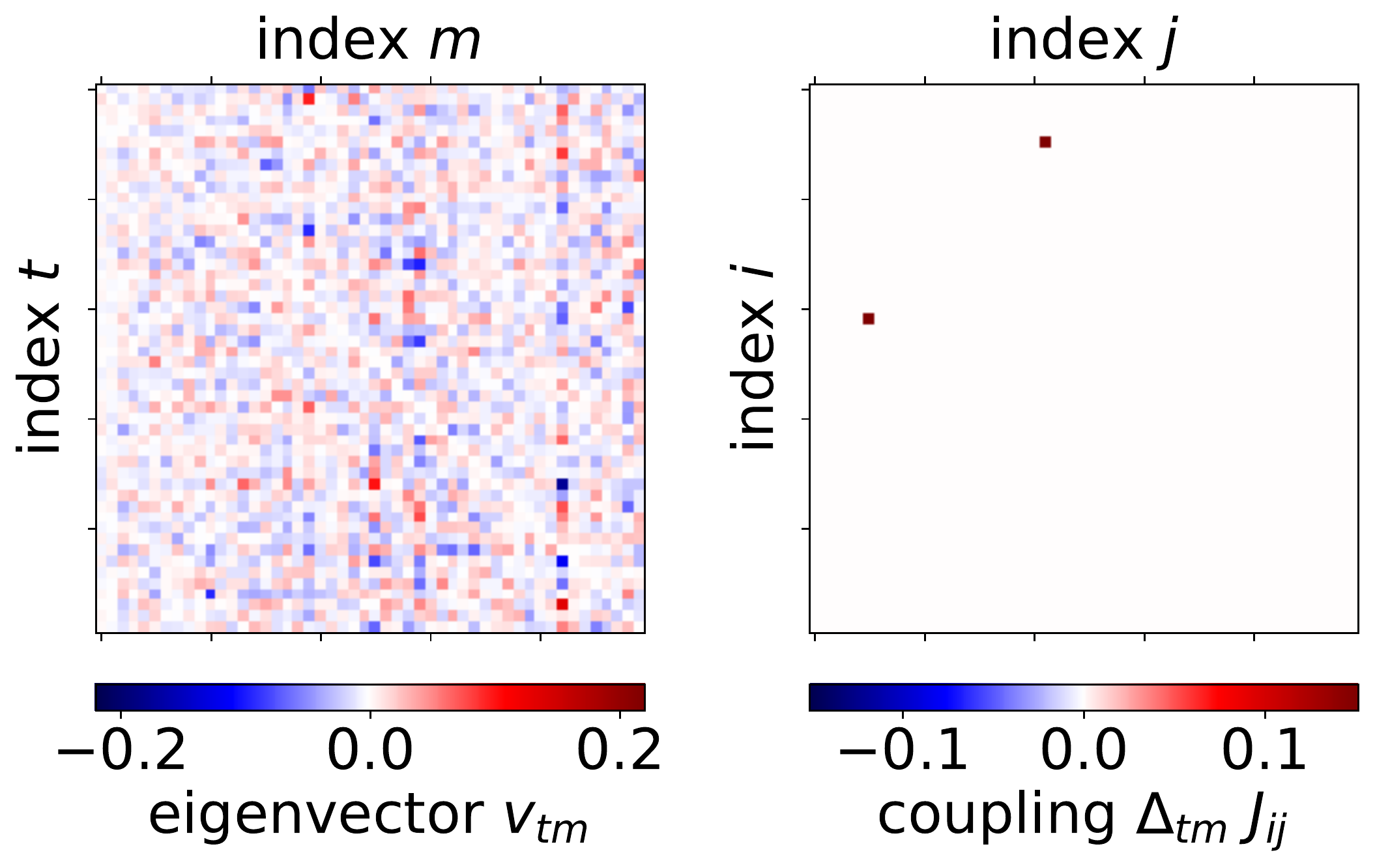}
	\caption{Example of simple coupling perturbation mapped to space of replacement rule. Compare with Figure~\ref{gr:comparison}C.}\label{si gr:transform ex}
\end{figure}

In physics experiments, it is possible to access directly quantities such as fields, couplings, and temperature to directly modify the Hamiltonian. With a phenomenological model of the neural network, however, we do not expect that perturbations protocols simply map to canonical perturbations. This means that for the natural perturbations we define in Eq~\ref{eq:replacement rule}, there is a corresponding representation in terms of the fields $\sethi$ and couplings $\setJij$, but it will generally be a complicated combination of many changes across the system (see Figures~\ref{gr:comparison} and \ref{si gr:transform ex}). In this sense, the pairwise maxent model serves as a representation of our inference process rather than a literal model of the neural network.

Besides the distinction between types of perturbations, we have a choice of which protocols to consider within each space. In the main text, we discuss individual neuron perturbations relative to neighbors as a way of approaching closed-loop control. However, other experimental protocols might be of interest such as probabilistic clamping to a fixed external reference frame always increasing, decreasing, or staying flat. An infinite variety of such variations are possible, and it may be the case that other representations turn out to be more appropriate for capturing simple experimental interventions. Though a complete basis can always, in principle, be transformed from one set of perturbations to another, such transformations may be difficult to perform accurately from limited and potentially noisy data extracted from experiments.

\begin{figure}\centering
	\includegraphics[width=.8\linewidth]{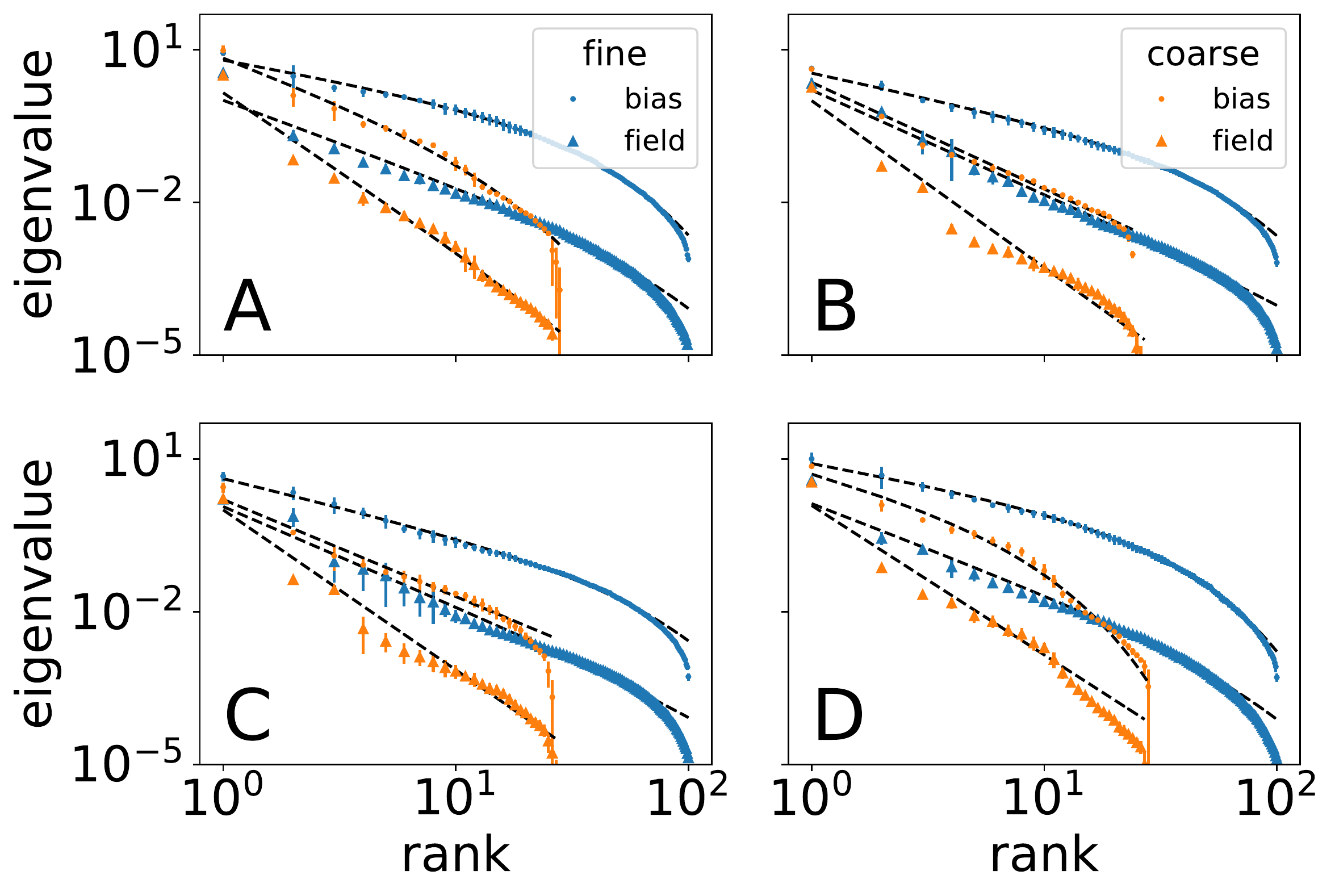}
	\caption{FIM eigenvalue spectra from field and bias perturbations for experiment 170419. We compare spectra for $\phi_{\rm fine}$ with $\phi_{\rm coarse}$. Panel letters indicate different neuron subsets of $N=50$ and averages over Monte Carlo samples with error bars representing standard deviations.}\label{si gr:field pert}
\end{figure}

\begin{figure}\centering
	\includegraphics[width=.5\linewidth]{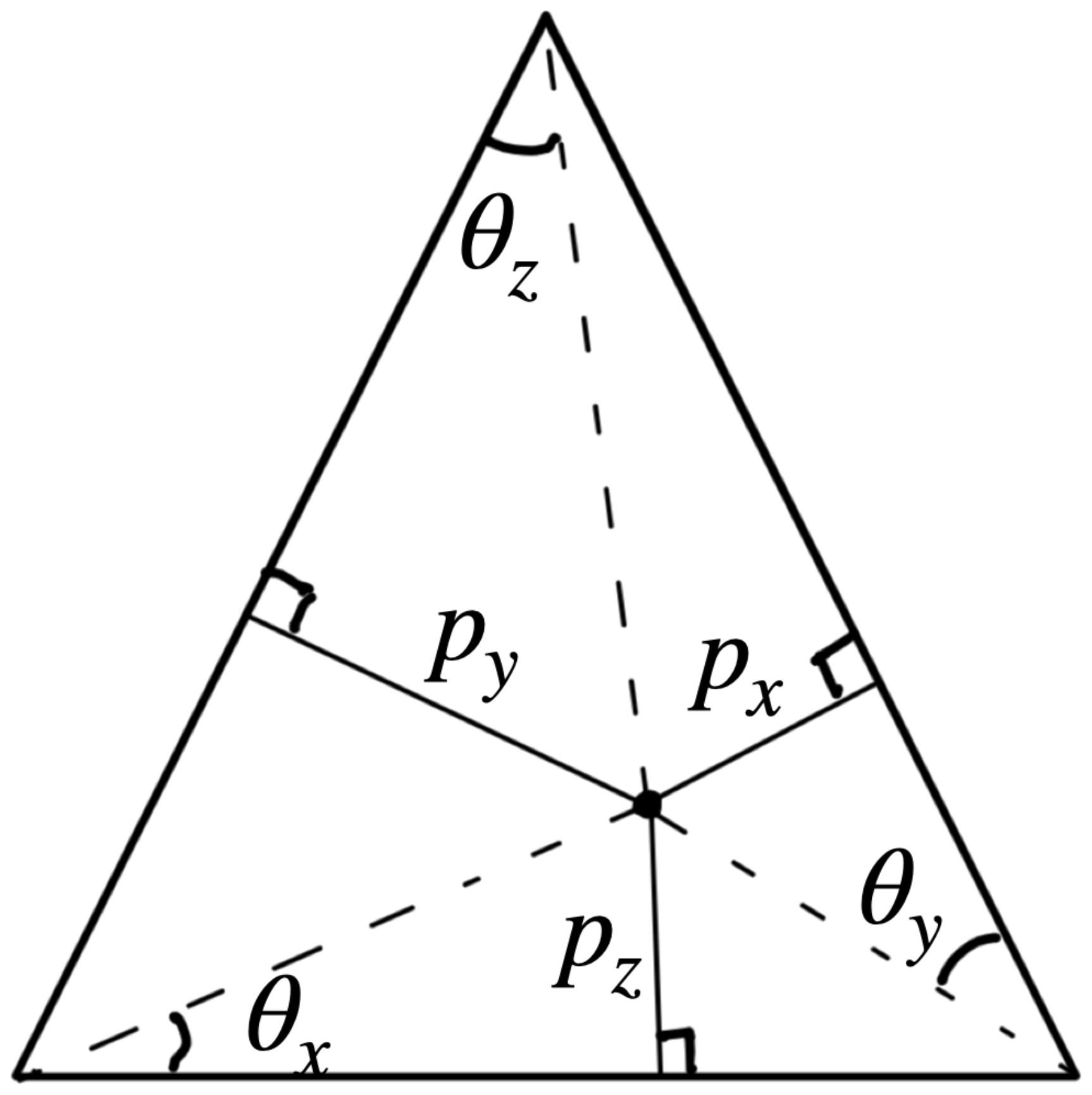}
	\caption{Probability simplex for three-state neuron with probabilities of being in each of three states denoted by $p_x$, $p_y$, and $p_z$ such that $p_x+p_y+p_z=1$.}\label{si gr:simplex}
\end{figure}

For the sake of completeness, we additionally compute the FIM for the example of clamping each neuron to an external reference frame---akin to clamping to an imaginary neighbor that is always in the up, down, or flat state. For the case of $K>2$ possible states of the neuron, clamping a neuron to a single state involves specifying the relative chance in the probabilities that the neuron is in the remaining $K-1$ states. We consider the case where clamping the neuron to a particular state $y$ changes the probabilities of the remaining two configurations in a way that traces out the geodesic towards $p_y\equiv p(s_m=y)=1$ in the two-dimensional simplex (Figure~\ref{si gr:simplex}). Experimental results may suggest more empirically grounded way of accounting for the relative change in probabilities. Taking this formulation, the local perturbation for the matcher neuron changes its bias as
\begin{align}
\begin{aligned}
	\tilde p_x &= p_x - \epsilon (1-p_y) \frac{\cos(\pi/2-\theta_y)}{\cos(\theta_y-\pi/6)}, \\
	\tilde p_y &= (1-\epsilon) p_y	 + \epsilon,\\
	\tilde p_z &= p_z - \epsilon (1-p_y) \frac{\cos(\pi/6+\theta_y)}{\cos(\theta_y-\pi/6)}.
\end{aligned}\label{si eq:dp}
\end{align}
In Figure~\ref{si gr:field pert}, we show a summary of results from such a protocol on the pairwise maxent model. We find that eigenvalues decay faster with rank when considering observable perturbations compared with natural perturbations for both fine and coarse collective synchrony. Across the maxent models we consider and the fine and coarse measure of synchrony, we consistently find that the decay exponent is faster for natural perturbations.

\section{Relating observable and canonical perturbations}\label{si sec:obs and can}
In the main text, we primarily focus on perturbations to the observed correlations, but we also find consistent differences between observable and natural perturbations. As we show in Figure~\ref{gr:eigenspectrum}, the spectrum tends to decay faster for canonical perturbations and they are smaller in general. To gain some insight into these differences, we consider the two types of perturbations for a highly simplified case of a binary neuron characterized by a mean activity level.

Consider states $s$ with probabilities $p(s)=\exp[-E(s)]/Z$ coarse-grained into a distribution $\phi(k) = \sum_{|s|=k} p(s)$, where the notation $|s|=k$ means that there are $k$ spins in the majority. Under the replacement rule, the probability of neuron $i$ being in one configuration $p_i$ is modified to become
\begin{align}
\begin{aligned}
	\tilde p_i &= (1-\epsilon)p_i + \epsilon\\
		&= p_i + (1-p)\epsilon.
\end{aligned}
\end{align}
Note that the derivative that returns the appropriate is of the form $-d/d[\log(1-p_i)]=(1-p_i)d/dp_i$. This would correspond to our observable perturbation, whereas the canonical perturbation would be with respect to the Langrangian multipliers, the fields and couplings for the pairwise maxent model.

The quantity of interest is the ``score function,'' or the expectation value of the second derivative of the probability distribution that gives us the entries of the FIM,
\begin{align}
	-\br{\frac{\partial^2\log\phi}{\partial [\log (1-p_i)]^2}}.\label{si eq:score}
\end{align}
The terms involving $\log\phi$ are of course the susceptibilities of the $k$th correlation function with the spin $i$. Since $\phi$ is a coarse-grained version of $p(s)$, we expect that the derivatives will reduces to linear combinations of correlation functions.

To simplify notation, we take the term inside the brackets having defined $y_i\equiv\log(1-p_i)$ to obtain
\begin{align}\begin{aligned}
	\frac{\partial^2\log\phi}{\partial y_i^2} &= \frac{\partial}{\partial y_i}\left[\frac{\partial h_i}{\partial y_i}\frac{\partial\log\phi}{\partial h_i}\right]\\
		&= \frac{\partial^2 h_i}{\partial y_i^2} \frac{\partial \log\phi}{\partial h_i} + \left[ \frac{\partial h_i}{\partial y_i} \right]^2 \frac{\partial^2 \log\phi}{\partial h_i^2},
\end{aligned}\label{eq:total derivative}
\end{align}
where we have introduced the field $h_i$ for neuron $i$. By taking the derivative with respect to the field, we have incurred a term proportional to the curvature in the change of variables as well as the square of the jacobian bringing us from $y_i$ to $h_i$.

Now, we calculate the jacobian terms. Note that we are only considering the domain where the relationship between $\log(1-p_i)$ and $h_i$ is analytic, otherwise we would have to deal with branch cuts. Given this caveat, we differentiate
\begin{align}
	\frac{\partial y_i}{\partial h_i} &= \frac{1}{p_i-1}\frac{\partial p_i}{\partial h_i}.
\intertext{Using the fact that $p_i = (\br{s_i}+1)/2$ and that the derivative is the susceptibility, or the variance of spin $i$, we obtain}
		&= \frac{1}{\br{s_i}-1} \left(1-\br{s_i}^2\right)\\
		&= -(1+\br{s_i}).
\end{align}
Then, the second derivative is
\begin{align}
	\frac{\partial^2 y_i}{\partial h_i^2} &= -\left(1-\br{s_i}^2\right).
\end{align}
Thus, the maxent model allows us to explicitly calculate the jacobian in terms of linear response quantities that tell us how the observables change under a small perturbation of the fields. 

Now, let us go back to the score function in Eq~\ref{si eq:score}. Note that because all the above quantities are already ensemble averages, we can factor them out of the brackets in Eq~\ref{si eq:score}. This means that we have
\begin{align}
	\frac{1}{\br{s_i}+1}\left[ \frac{1}{\br{s_i}-1}\br{\frac{\partial \log\phi}{\partial h_i}} -\frac{1}{\br{s_i}+1} \br{\frac{\partial^2\log\phi}{\partial h_i^2}} \right].
\end{align}
In contrast, if we were to do the same thing but take the derivative with respect to the fields in Eq~\ref{eq:total derivative} instead of the natural parameter, we would simply swap the $y_i$'s with the $h_i$'s. This means that we have the reciprocal of the jacobians,
\begin{align}
	(\br{s_i}+1)\left[ (\br{s_i}-1)\br{\frac{\partial \log\phi}{\partial y_i}} -(\br{s_i}+1) \br{\frac{\partial^2\log\phi}{\partial y_i^2}} \right].
\end{align}
In other words, changing variables gives us a different factor that depends on the mean magnetization of the entries of the FIM. This means that we may expect the overall eigenvalue spectrum to be, at the least, scaled differently. From numerical calculations, we find that the spectrum for observable perturbations to almost always be strictly greater than that for natural perturbations, which may reflect a change in variables.

\begin{figure}[p]\centering
	\includegraphics[width=.9\linewidth]{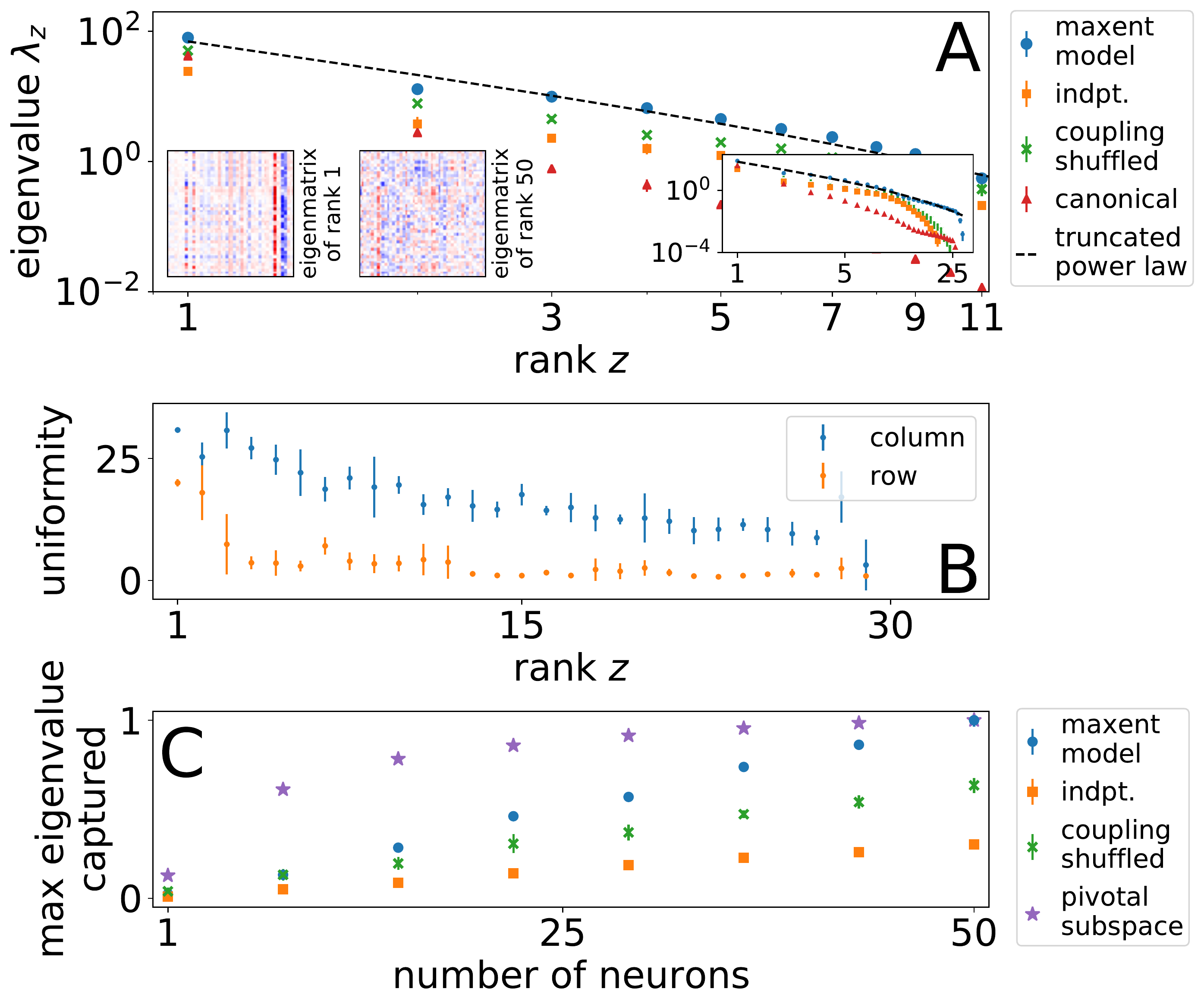}
	\caption{FIM overview for worm experiment 170419 with $\phi_{\rm coarse}$ instead of $\phi_{\rm fine}$ as shown in main text in Figure~\ref{gr:eigenspectrum}.}\label{si gr:eigenspectrum coarse 170419}
\end{figure}
\clearpage

\begin{figure}[p!]\centering
	\includegraphics[width=.9\linewidth]{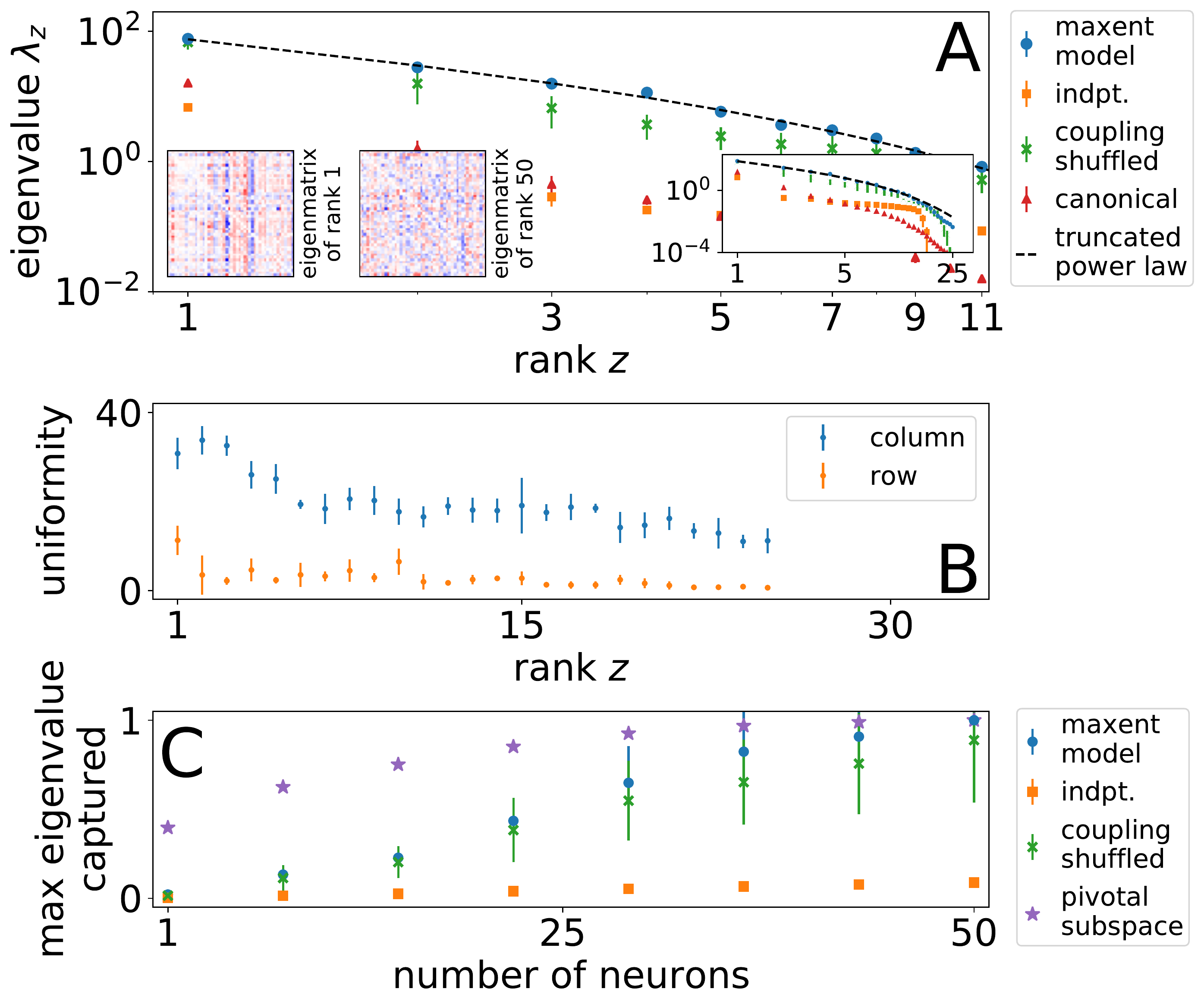}
	\caption{FIM overview for worm experiment 154139 with $\phi_{\rm coarse}$ instead of $\phi_{\rm fine}$ as shown in main text in Figure~\ref{gr:eigenspectrum}.}\label{si gr:eigenspectrum coarse 154139}
\end{figure}
\clearpage

\section{Data sets and samples}
We use the \celegans\ worm data from reference \cite{scholzPredictingNatural2018} and the same inverse maxent procedures detailed in reference \cite{chenSearchingCollective2019}. We analyze the two immobilized worms labeled with experiment numbers 170419 and 154139. For each worm, we consider a limited number of different subsets of $N=50$ neurons (about the maximum number of neurons that can be reliable measured from the data), where the neurons have been randomly selected amongst those that have visited all three states at least once.

\section{Experimental implementation}\label{si sec:experiment}
A realization of our thought experiment depends on developments in simultaneous use of recording, perturbation, and computational analysis. The experiment would require tracking time-averaged neural statistics before and during perturbation with the ability to extract simultaneously the discrete state of recorded neurons, a procedure that currently relies on post-experimental analysis to handle noise and changing fluorescence \cite{chenSearchingCollective2019}. Furthermore, single-neuron tracking is difficult and may be feasible with immobilized worms, but presents a challenge to perform accurately with freely moving ones. Furthermore, it is essential that the nature of the perturbation on membrane voltage be precisely calibrated in order to replicate as closely as possible theoretical clamping, which may require some characterization of the particular experimental techniques used. This presents an abbreviated list of some of the experimental difficulties that we foresee for implementing such an experiment.

Though we assume that the perturbation randomly ``flips'' the neural configuration of the focus neuron with some small probability at any given moment in time, the perturbation specified in Eq~\ref{eq:replacement rule} is compatible with variations of the experiment that may be more natural. For example, it may be more natural to force neurons up after a flat state but not directly from a down state. More specifically, it may be important to keep the slope of underlying neural activity from diverging. While such history-dependent clamping is compatible with our formulation that relies only on time averages, a rigorous comparison of techniques might reveal some protocols preferable over others in terms of feasibility or correspondence to theoretical predictions.

Finally, our proposed thought experiment generalizes beyond neural activity or the \celegans\ model system, where it may be possible to simultaneously observe system configuration while imposing a perturbation. With optogenetic tools, it may be interesting to consider other neural systems such as {\it in vitro} cortical cultures or the neurons involved in {\it Drosophila} flight.

\begin{figure}\centering
	\includegraphics[width=.9\linewidth]{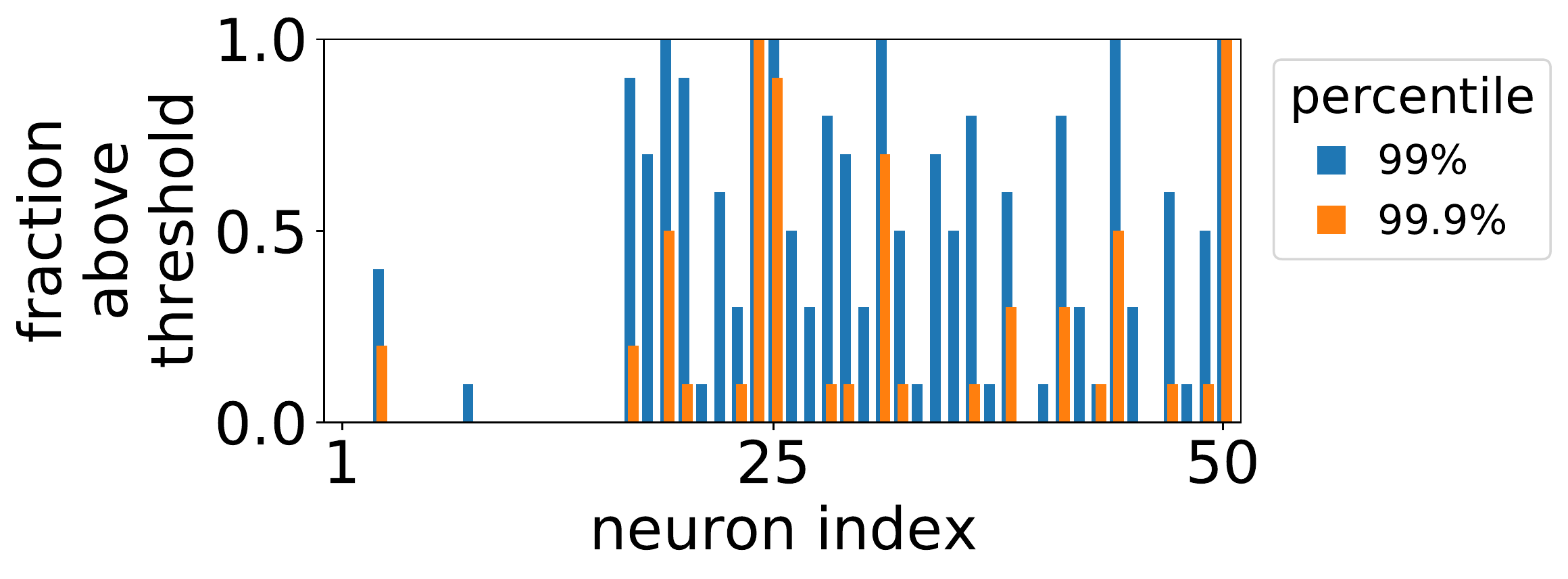}
	\caption{Frequency of repeated pivotal neurons from experiment 154139. See Figure~\ref{gr:piv neurons} for more details.}\label{si gr:piv neurons}
\end{figure}

\begin{figure}\centering
	\includegraphics[width=.9\linewidth]{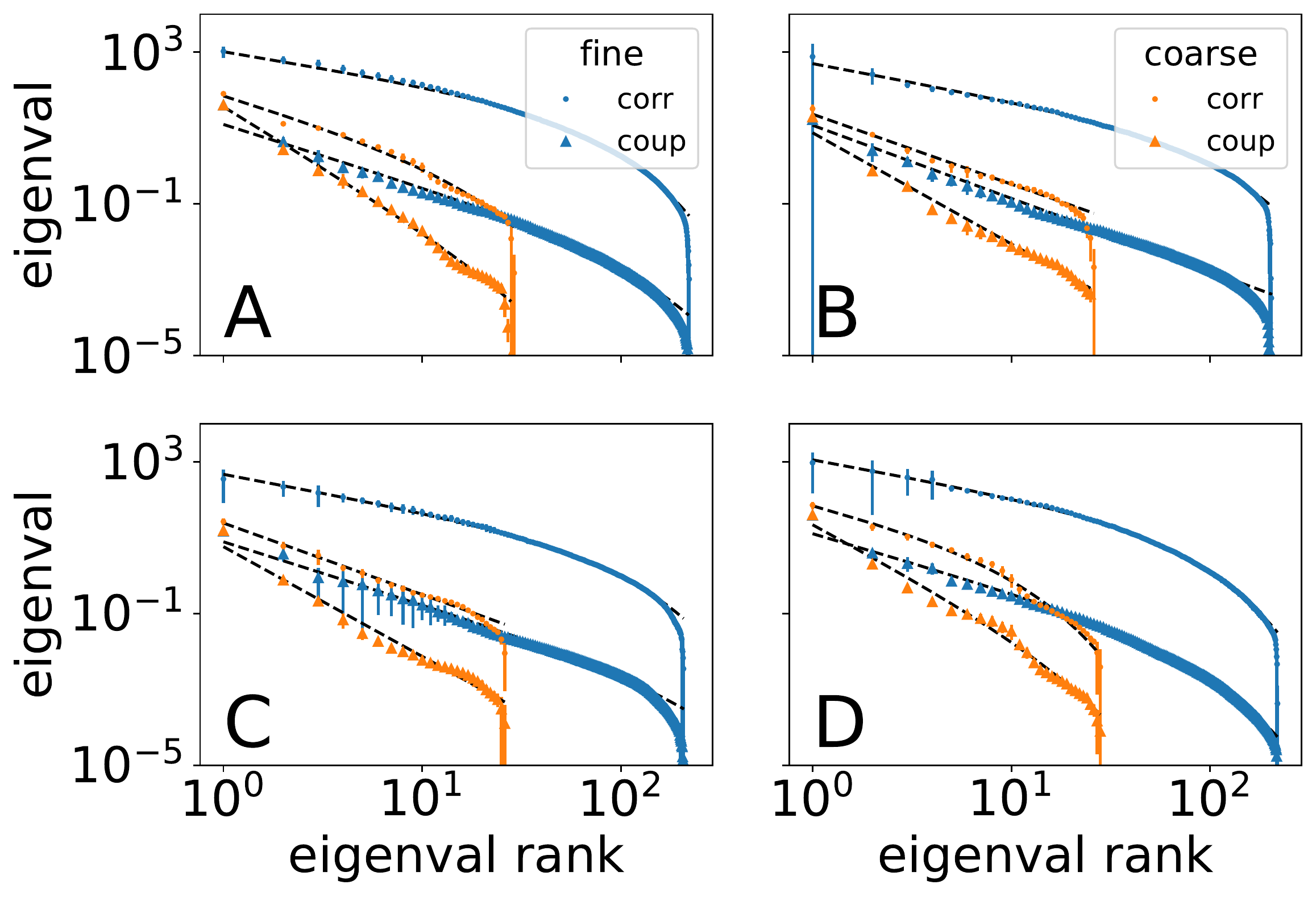}
	\caption{Power law with truncated exponential fit as dashed black lines to mean eigenvalue spectra for experiment 170419. Sensitivity for $\phi_{\rm fine}$ in blue and $\phi_{\rm coarse}$ in orange. Error bars show standard deviation over 10 Monte Carlo samples used to calculate FIM over 4 different subsets of neurons shown in each panel.}\label{si gr:spectra fit}
\end{figure}

\begin{figure}\centering
	\includegraphics[width=.9\linewidth]{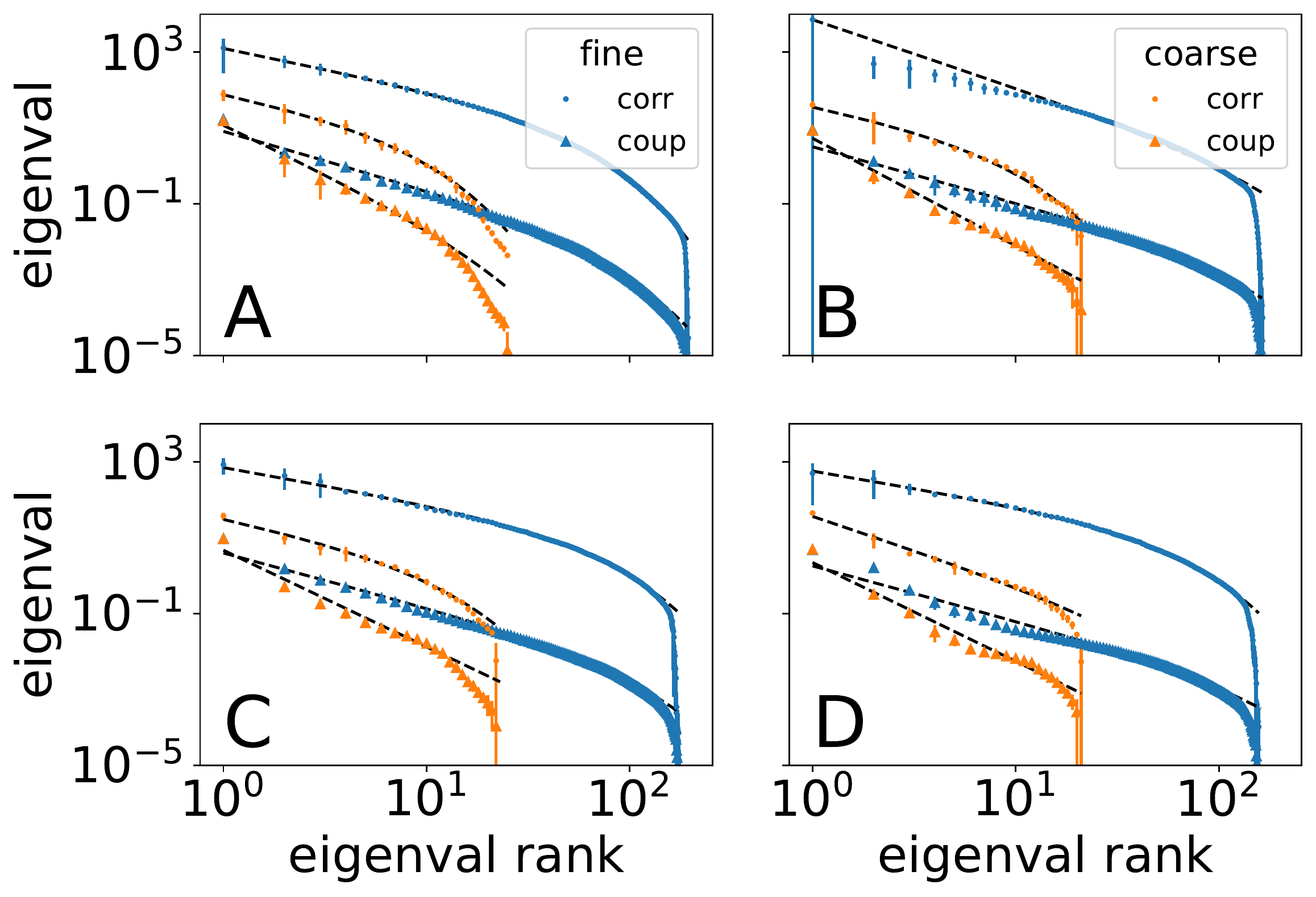}
	\caption{Power law with truncated exponential fit to mean eigenvalue spectra for experiment 154139. See Figure~\ref{si gr:spectra fit} for more details.}\label{si gr:spectra fit 2}
\end{figure}

%

%

\begin{thebibliography}{10}

\bibitem{gohHumanDisease2007}
Goh KI, Cusick ME, Valle D, Childs B, Vidal M, Barabasi AL.
\newblock The Human Disease Network.
\newblock Proc Natl Acad Sci USA. 2007;104(21):8685--8690.
\newblock doi:{10.1073/pnas.0701361104}.

\bibitem{vidalInteractomeNetworks2011}
Vidal M, Cusick ME, Barab{\'a}si AL.
\newblock Interactome {{Networks}} and {{Human Disease}}.
\newblock Cell. 2011;144:986--998.

\bibitem{zhangIntegratedSystems2013}
Zhang B, Gaiteri C, Bodea LG, Wang Z, McElwee J, Podtelezhnikov AA, et~al.
\newblock Integrated {{Systems Approach Identifies Genetic Nodes}} and
  {{Networks}} in {{Late}}-{{Onset Alzheimer}}'s {{Disease}}.
\newblock Cell. 2013;153(3):707--720.
\newblock doi:{10.1016/j.cell.2013.03.030}.

\bibitem{scholzPredictingNatural2018}
Scholz M, Linder AN, Randi F, Sharma AK, Yu X, Shaevitz JW, et~al.
\newblock Predicting Natural Behavior from Whole-Brain Neural Dynamics.
\newblock {Neuroscience}; 2018.

\bibitem{moroneSymmetryGroup2019}
Morone F, Makse HA.
\newblock Symmetry Group Factorization Reveals the Structure-Function Relation
  in the Neural Connectome of {{Caenorhabditis}} Elegans.
\newblock Nat Commun. 2019;10(1):4961.
\newblock doi:{10.1038/s41467-019-12675-8}.

\bibitem{liuControllabilityComplex2011}
Liu YY, Slotine JJ, Barab{\'a}si AL.
\newblock Controllability of Complex Networks.
\newblock Nature. 2011;473(7346):167--173.
\newblock doi:{10.1038/nature10011}.

\bibitem{tangIdentifyingControlling2012}
Tang Y, Gao H, Zou W, Kurths J.
\newblock Identifying {{Controlling Nodes}} in {{Neuronal Networks}} in
  {{Different Scales}}.
\newblock PLoS ONE. 2012;7(7):e41375.
\newblock doi:{10.1371/journal.pone.0041375}.

\bibitem{houwelingBehaviouralReport2008}
Houweling AR, Brecht M.
\newblock Behavioural report of single neuron stimulation in somatosensory
  cortex.
\newblock Nature. 2008;451(7174):65--68.

\bibitem{huberSparseOptical2008}
Huber D, Petreanu L, Ghitani N, Ranade S, Hrom{\'a}dka T, Mainen Z, et~al.
\newblock Sparse optical microstimulation in barrel cortex drives learned
  behaviour in freely moving mice.
\newblock Nature. 2008;451(7174):61--64.

\bibitem{katoGlobalDynamicsCelegans2015}
Kato S, Kaplan HS, Schr{\"o}del T, Skora S, Lindsay TH, Yemini E, et~al.
\newblock Global brain dynamics embed the motor command sequence of
  Caenorhabditis elegans.
\newblock Cell. 2015;163(3):656--669.

\bibitem{susoyCelegansMating2020}
Susoy V, Hung W, Witvliet D, Whitener JE, Wu M, Graham BJ, et~al.
\newblock Natural sensory context drives diverse brain-wide activity during C.
  elegans mating.
\newblock bioRxiv. 2020;doi:{10.1101/2020.09.09.289454}.

\bibitem{yanNetworkControl2017}
Yan G, V{\'e}rtes PE, Towlson EK, Chew YL, Walker DS, Schafer WR, et~al.
\newblock Network Control Principles Predict Neuron Function in the
  {{Caenorhabditis}} Elegans Connectome.
\newblock Nature. 2017;550(7677):519--523.
\newblock doi:{10.1038/nature24056}.

\bibitem{BorDan21}
Borriello E, Daniels BC.
\newblock {The basis of easy controllability in Boolean networks}.
\newblock in review. 2021; p. arXiv:2010.12075.

\bibitem{moroneFibrationSymmetries2020}
Morone F, Leifer I, Makse HA.
\newblock Fibration Symmetries Uncover the Building Blocks of Biological
  Networks.
\newblock Proc Natl Acad Sci USA. 2020;117(15):8306--8314.
\newblock doi:{10.1073/pnas.1914628117}.

\bibitem{ZanYanAlb17}
Za{\~{n}}udo JGT, Yang G, Albert R, Levine H.
\newblock {Structure-based control of complex networks with nonlinear
  dynamics}.
\newblock Proceedings of the National Academy of Sciences of the United States
  of America. 2017;114(28):7234--7239.
\newblock doi:{10.1073/pnas.1617387114}.

\bibitem{delferraroFindingInfluential2018}
Del~Ferraro G, Moreno A, Min B, Morone F, {P{\'e}rez-Ram{\'i}rez} {\'U},
  {P{\'e}rez-Cervera} L, et~al.
\newblock Finding Influential Nodes for Integration in Brain Networks Using
  Optimal Percolation Theory.
\newblock Nat Commun. 2018;9(1):2274.
\newblock doi:{10.1038/s41467-018-04718-3}.

\bibitem{lynnPhysicsBrain2019}
Lynn CW, Bassett DS.
\newblock The Physics of Brain Network Structure, Function, and Control.
\newblock Nat Rev Phys. 2019;1(5):318--332.
\newblock doi:{10.1038/s42254-019-0040-8}.

\bibitem{boydenMillisecondTimescale2005}
Boyden ES, Zhang F, Bamberg E, Nagel G, Deisseroth K.
\newblock Millisecond-timescale, genetically targeted optical control of neural
  activity.
\newblock Nature neuroscience. 2005;8(9):1263--1268.

\bibitem{mardinlyPreciseMultimodal2018}
Mardinly AR, Oldenburg IA, P{\'e}gard NC, Sridharan S, Lyall EH, Chesnov K,
  et~al.
\newblock Precise multimodal optical control of neural ensemble activity.
\newblock Nature neuroscience. 2018;21(6):881--893.

\bibitem{grosenickClosedLoop2015}
Grosenick L, Marshel JH, Deisseroth K.
\newblock Closed-loop and activity-guided optogenetic control.
\newblock Neuron. 2015;86(1):106--139.

\bibitem{zrennerClosedLoop2016}
Zrenner C, Belardinelli P, M{\"u}ller-Dahlhaus F, Ziemann U.
\newblock Closed-loop neuroscience and non-invasive brain stimulation: a tale
  of two loops.
\newblock Frontiers in cellular neuroscience. 2016;10:92.

\bibitem{kimIntegrationOptogenetics2017}
Kim CK, Adhikari A, Deisseroth K.
\newblock Integration of Optogenetics with Complementary Methodologies in
  Systems Neuroscience.
\newblock Nat Rev Neurosci. 2017;18(4):222--235.
\newblock doi:{10.1038/nrn.2017.15}.

\bibitem{pokalaInducibleTitratable2014}
Pokala N, Liu Q, Gordus A, Bargmann CI.
\newblock Inducible and Titratable Silencing of {{Caenorhabditis}} Elegans
  Neurons in Vivo with Histamine-Gated Chloride Channels.
\newblock Proc Natl Acad Sci USA. 2014;111(7):2770--2775.
\newblock doi:{10.1073/pnas.1400615111}.

\bibitem{gutrufFullyImplantable2018}
Gutruf P, Krishnamurthi V, V{\'a}zquez-Guardado A, Xie Z, Banks A, Su CJ,
  et~al.
\newblock Fully implantable optoelectronic systems for battery-free, multimodal
  operation in neuroscience research.
\newblock Nature Electronics. 2018;1(12):652--660.

\bibitem{barlowSingleUnits1972}
Barlow HB.
\newblock Single Units and Sensation: {{A}} Neuron Doctrine for Perceptual
  Psychology?
\newblock Perception. 1972;1:371--394.

\bibitem{quirogaSparseNot2007}
Quiroga RQ, Kreiman G, Koch C, Fried I.
\newblock Sparse but Not `{{Grandmother}}-Cell' Coding in the Medial Temporal
  Lobe.
\newblock Cell. 2007;12(3):87--91.

\bibitem{danielsSparseCode2012}
Daniels BC, Krakauer DC, Flack JC.
\newblock Sparse Code of Conflict in a Primate Society.
\newblock Proc Natl Acad Sci USA. 2012;109(35):14259--14264.
\newblock doi:{10.1073/pnas.1203021109}.

\bibitem{olshausenSparseCoding1997}
Olshausen BA, Field DJ.
\newblock Sparse Coding with an Overcomplete Basis Set: {{A}} Strategy Employed
  by {{V1}}?
\newblock Vision Research. 1997;37(23):3311--3325.
\newblock doi:{10.1016/S0042-6989(97)00169-7}.

\bibitem{spanneQuestioningRole2015}
Spanne A, J{\"o}rntell H.
\newblock Questioning the Role of Sparse Coding in the Brain.
\newblock Trends in Neurosciences. 2015;38(7):417--427.
\newblock doi:{10.1016/j.tins.2015.05.005}.

\bibitem{beyelerNeuralCorrelates2019}
Beyeler M, Rounds EL, Carlson KD, Dutt N, Krichmar JL.
\newblock Neural Correlates of Sparse Coding and Dimensionality Reduction.
\newblock PLoS Comput Biol. 2019;15(6):e1006908.
\newblock doi:{10.1371/journal.pcbi.1006908}.

\bibitem{oizumiInformationLoss2011}
Oizumi M, Okada M, Amari SI.
\newblock Information {{Loss Associated}} with {{Imperfect Observation}} and
  {{Mismatched Decoding}}.
\newblock Front Comput Neurosci. 2011;5.
\newblock doi:{10.3389/fncom.2011.00009}.

\bibitem{grayNavigationCelegans2005}
Gray JM, Hill JJ, Bargmann CI.
\newblock A circuit for navigation in Caenorhabditis elegans.
\newblock Proceedings of the National Academy of Sciences.
  2005;102(9):3184--3191.

\bibitem{ikedaContextdependentOperation2020}
Ikeda M, Nakano S, Giles AC, Xu L, Costa WS, Gottschalk A, et~al.
\newblock Context-Dependent Operation of Neural Circuits Underlies a Navigation
  Behavior in {{{\emph{Caenorhabditis}}}}{\emph{ Elegans}}.
\newblock Proc Natl Acad Sci USA. 2020;117(11):6178--6188.
\newblock doi:{10.1073/pnas.1918528117}.

\bibitem{ouelletteGateandSwitchModel2018}
Ouellette MH, Desrochers MJ, Gheta I, Ramos R, Hendricks M.
\newblock A {{Gate}}-and-{{Switch Model}} for {{Head Orientation Behaviors}} in
  {{{\emph{Caenorhabditis}}}}{\emph{ Elegans}}.
\newblock eNeuro. 2018;5(6):ENEURO.0121--18.2018.
\newblock doi:{10.1523/ENEURO.0121-18.2018}.

\bibitem{changHypoxiaHIF12008}
Chang AJ, Bargmann CI.
\newblock Hypoxia and the {{HIF}}-1 Transcriptional Pathway Reorganize a
  Neuronal Circuit for Oxygen-Dependent Behavior in {{Caenorhabditis}} Elegans.
\newblock Proc Natl Acad Sci USA. 2008;105(20):7321--7326.
\newblock doi:{10.1073/pnas.0802164105}.

\bibitem{lashleyBrainMechanisms1929}
Lashley KS.
\newblock Brain Mechanisms and Intelligence: {{A}} Quantitative Study of
  Injuries to the Brain.
\newblock Behavior Research Fund Monographs. {University of Chicago Press};
  1929.

\bibitem{damasloReturnPhineas1994}
Damaslo H, Grabowski T, Frank R, Galaburda AM, Damasio AR.
\newblock The {{Return}} of {{Phineas Gage}}: {{Clues About}} the {{Brain}}
  from the {{Skull}} of a {{Famous Patient}}.
\newblock Science. 1994;264:5.

\bibitem{stephensDimensionalityDynamics2008}
Stephens GJ, {Johnson-Kerner} B, Bialek W, Ryu WS.
\newblock Dimensionality and {{Dynamics}} in the {{Behavior}} of {{C}}.
  Elegans.
\newblock PLoS Comput Biol. 2008;4(4):e1000028.
\newblock doi:{10.1371/journal.pcbi.1000028}.

\bibitem{schneidmanWeakPairwise2006}
Schneidman E, Berry MJ, Segev R, Bialek W.
\newblock Weak Pairwise Correlations Imply Strongly Correlated Network States
  in a Neural Population.
\newblock Nature. 2006;440(7087):1007--1012.
\newblock doi:{10.1038/nature04701}.

\bibitem{hopfieldNeuralNetworks1982}
Hopfield JJ.
\newblock Neural Networks and Physical Systems with Emergent Collective
  Computational Abilities.
\newblock Proc Natl Acad Sci USA. 1982;79:2554--2558.

\bibitem{transtrumModelReduction2014}
Transtrum MK, Qiu P.
\newblock Model {{Reduction}} by {{Manifold Boundaries}}.
\newblock Phys Rev Lett. 2014;113(9):098701.
\newblock doi:{10.1103/PhysRevLett.113.098701}.

\bibitem{transtrumPerspectiveSloppiness2015}
Transtrum MK, Machta BB, Brown KS, Daniels BC, Myers CR, Sethna JP.
\newblock Perspective: {{Sloppiness}} and Emergent Theories in Physics,
  Biology, and Beyond.
\newblock J Chem Phys. 2015;143(1):010901.
\newblock doi:{10.1063/1.4923066}.

\bibitem{maunsellFunctionalProperties1983}
Maunsell JH, Van~Essen DC.
\newblock Functional Properties of Neurons in Middle Temporal Visual Area of
  the Macaque Monkey. {{I}}. {{Selectivity}} for Stimulus Direction, Speed, and
  Orientation.
\newblock Journal of Neurophysiology. 1983;49(5):1127--1147.
\newblock doi:{10.1152/jn.1983.49.5.1127}.

\bibitem{wuPopulationCoding2002}
Wu S, Amari Si, Nakahara H.
\newblock Population {{Coding}} and {{Decoding}} in a {{Neural Field}}: {{A
  Computational Study}}.
\newblock Neural Computation. 2002;14(5):999--1026.
\newblock doi:{10.1162/089976602753633367}.

\bibitem{nguyenWholeBrain2016}
Nguyen JP, Shipley FB, Linder AN, Plummer GS, Liu M, Setru SU, et~al.
\newblock Whole-brain calcium imaging with cellular resolution in freely
  behaving Caenorhabditis elegans.
\newblock Proceedings of the National Academy of Sciences.
  2016;113(8):E1074--E1081.

\bibitem{venkatachalamPanneuronalImaging2016}
Venkatachalam V, Ji N, Wang X, Clark C, Mitchell JK, Klein M, et~al.
\newblock Pan-Neuronal Imaging in Roaming {{{\emph{Caenorhabditis}}}}{\emph{
  Elegans}}.
\newblock Proc Natl Acad Sci USA. 2016;113(8):E1082--E1088.
\newblock doi:{10.1073/pnas.1507109113}.

\bibitem{stirmanRealTime2011}
Stirman JN, Crane MM, Husson SJ, Wabnig S, Schultheis C, Gottschalk A, et~al.
\newblock Real-time multimodal optical control of neurons and muscles in freely
  behaving Caenorhabditis elegans.
\newblock Nature methods. 2011;8(2):153--158.

\bibitem{bermanPredictabilityHierarchy2016}
Berman GJ, Bialek W, Shaevitz JW.
\newblock Predictability and Hierarchy in {{{\emph{Drosophila}}}} Behavior.
\newblock Proc Natl Acad Sci USA. 2016;113(42):11943--11948.
\newblock doi:{10.1073/pnas.1607601113}.

\bibitem{merchanSufficiencyPairwise2016}
Merchan L, Nemenman I.
\newblock On the {{Sufficiency}} of {{Pairwise Interactions}} in {{Maximum
  Entropy Models}} of {{Networks}}.
\newblock J Stat Phys. 2016;162(5):1294--1308.
\newblock doi:{10.1007/s10955-016-1456-5}.

\bibitem{chenSearchingCollective2019}
Chen X, Randi F, Leifer AM, Bialek W.
\newblock Searching for Collective Behavior in a Small Brain.
\newblock Phys Rev E. 2019;99(5):052418.
\newblock doi:{10.1103/PhysRevE.99.052418}.

\bibitem{xuHighlyEfficient2016}
Xu S, Chisholm AD.
\newblock Highly Efficient Optogenetic Cell Ablation in {{C}}. Elegans Using
  Membrane-Targeted {{miniSOG}}.
\newblock Sci Rep. 2016;6(1):21271.
\newblock doi:{10.1038/srep21271}.

\bibitem{sukClosedLoop2017}
Suk HJ, van Welie I, Kodandaramaiah SB, Allen B, Forest CR, Boyden ES.
\newblock Closed-loop real-time imaging enables fully automated cell-targeted
  patch-clamp neural recording in vivo.
\newblock Neuron. 2017;95(5):1037--1047.

\bibitem{newmanOptoFeedback2015}
Newman JP, Fong Mf, Millard DC, Whitmire CJ, Stanley GB, Potter SM.
\newblock Optogenetic feedback control of neural activity.
\newblock Elife. 2015;4:e07192.

\bibitem{prinzDynamicClamp2004}
Prinz AA, Abbott L, Marder E.
\newblock The dynamic clamp comes of age.
\newblock Trends in neurosciences. 2004;27(4):218--224.

\bibitem{morcosDirectcouplingAnalysis2011}
Morcos F, Pagnani A, Lunt B, Bertolino A, Marks DS, Sander C, et~al.
\newblock Direct-Coupling Analysis of Residue Coevolution Captures Native
  Contacts across Many Protein Families.
\newblock Proc Natl Acad Sci USA. 2011;108(49):E1293--E1301.

\bibitem{volkovInferringSpecies2009}
Volkov I, Banavar JR, Hubbell SP, Maritan A.
\newblock Inferring Species Interactions in Tropical Forests.
\newblock Proc Natl Acad Sci USA. 2009;106(33):13854--13859.
\newblock doi:{10.1073/pnas.0903244106}.

\bibitem{coverElementsInformation2006}
Cover TM, Thomas JA.
\newblock Elements of {{Information Theory}}.
\newblock 2nd ed. {Hoboken}: {John Wiley \& Sons}; 2006.

\bibitem{jaynesInformationTheory1957}
Jaynes ET.
\newblock Information {{Theory}} and {{Statistical Mechanics}}.
\newblock Phys Rev. 1957;106(4):620--630.
\newblock doi:{10.1103/PhysRev.106.620}.

\bibitem{leeStatisticalMechanics2015}
Lee ED, Broedersz CP, Bialek W.
\newblock Statistical {{Mechanics}} of the {{US Supreme Court}}.
\newblock J Stat Phys. 2015;160(2):275--301.
\newblock doi:{10.1007/s10955-015-1253-6}.

\bibitem{leeSensitivityCollective2020}
Lee ED, Katz DM, Bommarito~II MJ, Ginsparg PH.
\newblock Sensitivity of Collective Outcomes Identifies Pivotal Components.
\newblock J R Soc Interface. 2020;17(20190873).

\bibitem{tkacikSpinGlass2009}
Tkacik G, Schneidman E, Berry~II MJ, Bialek W.
\newblock Spin Glass Models for a Network of Real Neurons.
\newblock arXiv:09125409 [q-bio]. 2009;.

\bibitem{tkacikSearchingCollective2014}
Tka{\v c}ik G, Marre O, Amodei D, Schneidman E, Bialek W, Berry MJ.
\newblock Searching for {{Collective Behavior}} in a {{Large Network}} of
  {{Sensory Neurons}}.
\newblock PLoS Comput Biol. 2014;10(1):e1003408.
\newblock doi:{10.1371/journal.pcbi.1003408}.

\bibitem{meshulamCoarseGraining2019}
Meshulam L, Gauthier JL, Brody CD, Tank DW, Bialek W.
\newblock Coarse {{Graining}}, {{Fixed Points}}, and {{Scaling}} in a {{Large
  Population}} of {{Neurons}}.
\newblock Phys Rev Lett. 2019;123(17):178103.
\newblock doi:{10.1103/PhysRevLett.123.178103}.

\bibitem{bartonIsingModels2013}
Barton J, Cocco S.
\newblock Ising Models for Neural Activity Inferred via Selective Cluster
  Expansion: Structural and Coding Properties.
\newblock J Stat Mech. 2013;2013(03):P03002.
\newblock doi:{10.1088/1742-5468/2013/03/P03002}.

\bibitem{roudiPairwiseMaximum2009}
Roudi Y, Nirenberg S, Latham PE.
\newblock Pairwise {{Maximum Entropy Models}} for {{Studying Large Biological
  Systems}}: {{When They Can Work}} and {{When They Can}}'t.
\newblock PLoS Comput Biol. 2009;5(5):e1000380.
\newblock doi:{10.1371/journal.pcbi.1000380}.

\bibitem{amariInformationGeometry2016}
Amari Si.
\newblock Information Geometry and Its Applications. vol. 194 of Applied
  Mathematical Sciences.
\newblock {Springer Japan}; 2016.

\bibitem{bialekRediscoveringPower2007}
Bialek W, Ranganathan R.
\newblock Rediscovering the Power of Pairwise Interactions.
\newblock arXiv:07124397 [q-bio]. 2007;.

\bibitem{tkacikSimplestMaximum2013a}
Tka{\v c}ik G, Marre O, Mora T, Amodei D, Berry~II MJ, Bialek W.
\newblock The Simplest Maximum Entropy Model for Collective Behavior in a
  Neural Network.
\newblock J Stat Mech. 2013;2013(03):P03011.
\newblock doi:{10.1088/1742-5468/2013/03/P03011}.

\bibitem{bialekBiophysicsSearching2012}
Bialek WS.
\newblock Biophysics: Searching for Principles.
\newblock {Princeton, NJ}: {Princeton University Press}; 2012.

\bibitem{variance_note}
As a more intuitive formulation of Eq~\ref{eq:fim}, we could assign to each
  possible collective configuration a ``synchrony energy'' $\mathcal{E}$ such
  that its probability can be written $\phi \propto \exp[-\mathcal{E}]$. Such
  an effective energy represents a coarse-graining over the microscopic states
  corresponding to a collective configuration. Then, Eq~\ref{eq:fim} can also
  be written as the limiting quantity involving the change in energies
  $\Delta\mathcal{E}$ under such a perturbation, the variance
  $\lim_{\epsilon\rightarrow0} \left(\br{\Delta \mathcal{E}^2}-\br{\Delta
  \mathcal{E}}^2\right)/\epsilon^2$ \cite{leeSensitivityCollective2020}. This
  is a measure of how differently the log-probability of each collective
  configuration changes. Thus, the Fisher information is proportional to the
  variance of the effective energy such that more sensitive directions of
  change are ones that maximally scatter the collective distribution.;.

\bibitem{danielsDualCoding2017}
Daniels BC, Flack JC, Krakauer DC.
\newblock Dual {{Coding Theory Explains Biphasic Collective Computation}} in
  {{Neural Decision}}-{{Making}}.
\newblock Front Neurosci. 2017;11:313.
\newblock doi:{10.3389/fnins.2017.00313}.

\bibitem{mitraWmatrixGeometry1969}
Mitra D.
\newblock Wmatrix and the Geometry of Model Equivalence and Reduction.
\newblock Proc Inst Electr Eng UK. 1969;116(6):1101.
\newblock doi:{10.1049/piee.1969.0206}.

\bibitem{royRelationFIM2009}
Roy P, Cela A, Hamam Y.
\newblock On the Relation of {{FIM}} and {{Controllability Gramian}}.
\newblock In: 2009 {{IEEE International Symposium}} on {{Industrial Embedded
  Systems}}. {Lausanne, Switzerland}: {IEEE}; 2009. p. 37--41.

\bibitem{liuConvergenceFundamental2006}
Liu J, Elia N.
\newblock Convergence of {{Fundamental Limitations}} in {{Information}},
  {{Estimation}}, and {{Control}}.
\newblock In: Proceedings of the 45th {{IEEE Conference}} on {{Decision}} and
  {{Control}}. {San Diego, CA, USA}: {IEEE}; 2006. p. 5609--5614.

\bibitem{rajanEigenvalueSpectra2006}
Rajan K, Abbott LF.
\newblock Eigenvalue {{Spectra}} of {{Random Matrices}} for {{Neural
  Networks}}.
\newblock Phys Rev Lett. 2006;97(18):188104.
\newblock doi:{10.1103/PhysRevLett.97.188104}.

\bibitem{weigtIdentificationDirect2009}
Weigt M, White RA, Szurmant H, Hoch JA, Hwa T.
\newblock Identification of Direct Residue Contacts in Protein-Protein
  Interaction by Message Passing.
\newblock Proc Natl Acad Sci USA. 2009;106(1):67--72.
\newblock doi:{10.1073/pnas.0805923106}.

\bibitem{oliveriGlobalRegulatory2008}
Oliveri P, Tu Q, Davidson EH.
\newblock Global Regulatory Logic for Specification of an Embryonic Cell
  Lineage.
\newblock Proc Natl Acad Sci USA. 2008;105(16):5955--5962.
\newblock doi:{10.1073/pnas.0711220105}.

\bibitem{liEncodingRegulatory2014}
Li E, Cui M, Peter IS, Davidson EH.
\newblock Encoding Regulatory State Boundaries in the Pregastrular Oral
  Ectoderm of the Sea Urchin Embryo.
\newblock Proc Natl Acad Sci USA. 2014;111(10):E906--E913.
\newblock doi:{10.1073/pnas.1323105111}.

\bibitem{fim_footnote}
In calculating the Fisher information matrix, we use $\epsilon=10^{-4}$, but
  the results do not depend strongly at this range of perturbation size.
  Furthermore, this lower bound on the strength of perturbation is complemented
  from above by two additional bounds. First, it is unlikely that the system
  can relax back from a large perturbation. Second, a sufficiently large
  perturbation breaks the linear assumptions we make. These upper bounds might
  not separate with the lower bound for measurement, which would pose a problem
  for measurement, but this is a question that must be answered
  experimentally.;.

\bibitem{candesExactMatrix2009}
Cand{\`e}s EJ, Recht B.
\newblock Exact {{Matrix Completion}} via {{Convex Optimization}}.
\newblock Found Comput Math. 2009;9(6):717--772.
\newblock doi:{10.1007/s10208-009-9045-5}.

\bibitem{shannonMathematicalTheory1948}
Shannon CE.
\newblock A {{Mathematical Theory}} of {{Communication}}.
\newblock Bell Syst Tech J. 1948;27:379--423, 623--656.

\bibitem{boltzmann_note}
In the case of a Boltzmann-type model, this perturbative term can be expressed
  as a change in the energy of state $Y$ relative to the average change in
  energy. For the pairwise maxent model, the energy is a sum over the set of
  all couplings, not necessarily only the ones associated with the matcher
  neurons and its neighbors. If the structure of the coupling network were
  locally tree-like such that local neighbors interacted more strongly with the
  matcher than ones further away, then the change in energy could be
  approximated as a sum only over the couplings connecting the matcher with its
  neighbors. This is not generally the case for the networks we consider in the
  data.;.

\bibitem{brennerAdaptiveRescaling2000}
Brenner N, Bialek W, {de Ruyter van Steveninck} R.
\newblock Adaptive {{Rescaling Maximizes Information Transmission}}.
\newblock Neuron. 2000;26(3):695--702.
\newblock doi:{10.1016/S0896-6273(00)81205-2}.

\bibitem{transtrumGeometryNonlinear2011}
Transtrum MK, Machta BB, Sethna JP.
\newblock Geometry of Nonlinear Least Squares with Applications to Sloppy
  Models and Optimization.
\newblock Phys Rev E. 2011;83(3):036701.
\newblock doi:{10.1103/PhysRevE.83.036701}.

\bibitem{transtrumWhyAre2010}
Transtrum MK, Machta BB, Sethna JP.
\newblock Why Are {{Nonlinear Fits}} to {{Data}} so {{Challenging}}?
\newblock Phys Rev Lett. 2010;104(6):060201.
\newblock doi:{10.1103/PhysRevLett.104.060201}.

\bibitem{machtaParameterSpace2013}
Machta BB, Chachra R, Transtrum MK, Sethna JP.
\newblock Parameter {{Space Compression Underlies Emergent Theories}} and
  {{Predictive Models}}.
\newblock Science. 2013;342(6158):604--607.
\newblock doi:{10.1126/science.1238723}.

\bibitem{danielsParameterEstimation2018}
Daniels BC, Dobrzy{\'n}ski M, Fey D.
\newblock Parameter {{Estimation}}, {{Sloppiness}}, and {{Model
  Identifiability}}.
\newblock In: Munsky B, editor. Quantitative Biology: Theory, Computational
  Methods, and Models. {Cambridge, Massachusetts}: {The MIT Press}; 2018. p.
  271--292.

\bibitem{nguyenInverseStatistical2017}
Nguyen HC, Zecchina R, Berg J.
\newblock Inverse Statistical Problems: From the Inverse {{Ising}} Problem to
  Data Science.
\newblock Advances in Physics. 2017;66(3):197--261.
\newblock doi:{10.1080/00018732.2017.1341604}.

\bibitem{leeConvenientInterface2019}
Lee ED, Daniels BC.
\newblock Convenient {{Interface}} to {{Inverse Ising}} ({{ConIII}}): {{A
  Python}} 3 {{Package}} for {{Solving Ising}}-{{Type Maximum Entropy Models}}.
\newblock JORS. 2019;7(1):3.
\newblock doi:{10.5334/jors.217}.

\bibitem{dudikPerformanceGaurantees2004}
Dudik M, Phillips SJ, Schapire RE.
\newblock Performance guarantees for regularized maximum entropy density
  estimation.
\newblock In: International Conference on Computational Learning Theory.
  Springer; 2004. p. 472--486.

\bibitem{broderickFasterSolutions2007}
Broderick T, Dudik M, Tkacik G, Schapire RE, Bialek W.
\newblock Faster Solutions of the Inverse Pairwise {{Ising}} Problem.
\newblock arXiv:07122437 [cond-mat, q-bio]. 2007;.

\end{thebibliography}

\end{document}